\documentclass[%
  twoside,
  reprint,
  amsmath,amssymb,
  aps,
  pra,
  nofootinbib,
  showpacs,
  superscriptaddress,
  a4paper
]{revtex4-1}

\usepackage{graphicx}%
\usepackage[usenames,dvipsnames]{xcolor}
\usepackage{siunitx}
\usepackage{enumitem}

\usepackage{newfloat}
\DeclareFloatingEnvironment[
  name = {3D FIG.}
]{tdfigure}
\usepackage[caption=false]{subfig}
\newcounter{subtdfigure}
\newcounter{subtdfigure@save}

\usepackage{tabularx}
\usepackage{booktabs}
\usepackage{multirow}

\usepackage{titlesec}

\usepackage{array}

\usepackage[utf8]{inputenc}
\usepackage[T1]{fontenc}

\usepackage{lipsum}

\graphicspath{{Figures/}{}}

\usepackage[
centering, includefoot,
text={7.1in,10.2in},
total={6.3in,8.75in},
top=0.8in, left=0.62in,
]{geometry}

\usepackage[
  bookmarks=false,
  colorlinks,
  linkcolor=blue,
  urlcolor=blue,
  citecolor=blue,
  plainpages=false,
  pdfpagelabels,
  final,
  breaklinks=true
]{hyperref}
\hypersetup{
pdftitle={The imaginary part of the high-harmonic cutoff}, 
pdfauthor={Emilio Pisanty, Marcelo F Ciappina and Maciej Lewenstein}
}

\usepackage{xmpincl}
\includexmp{metadata}

\usepackage{natbib}
\makeatletter \def\NAT@def@citea{\def\@citea{\NAT@separator\,}} \makeatother
\newcommand{\citer}[1]{Ref.~\citealp{#1}}

\newcommand{\reffig}[1]{Fig.~\ref{#1}}

\newcommand{\reffigtd}[1]{3D Fig.~\ref{#1}}
\newcommand{\reffigtdsub}[2]{3D Fig.~\ref{#1}\ref{#2}}

\newcommand{\citenisteq}[1]{\cite[Eq.\,(\href{http://dlmf.nist.gov/#1}{#1})]{NIST_handbook}}

\newcommand{\orcid}[1]{%
  \href{%
    https://orcid.org/#1%
  }{%
   \,\protect\includegraphics[width=8pt]{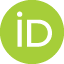}%
  }%
}

\renewcommand{\d}{\ensuremath{\textrm{d}}}
\renewcommand{\Re}{\operatorname{Re}}
\renewcommand{\Im}{\operatorname{Im}}
\DeclareMathOperator{\Ai}{Ai}
\usepackage{physics}

  \newcommand{\vbp}{\vb{p}}
  \newcommand{\vbk}{\vb{k}}
  \newcommand{\vba}{\vb{A}}
  \newcommand{\vbf}{\vb{F}}
  \newcommand{\vbd}{\vb{d}}
  \newcommand{\vbD}{\vb{D}}
  \newcommand{\vbv}{\vb{v}}

\newcommand{\vbalpha}{\boldsymbol{\alpha}}

\newcommand{\ue}[1]{\hat{\vb{e}}_{#1}}

\newcommand{\hc}{\ensuremath{\mathrm{hc}}}
\newcommand{\sep}{\ensuremath{\mathrm{sep}}}

\newcommand{\cc}{\ensuremath{\mathrm{c.c.}}}

\newcommand{\Y}{\Upsilon}

\DeclareSIUnit{\au}{{a.u.}}

\newcommand{\nhphantom}[1]{\sbox0{#1}\hspace{-\the\wd0}}

\begin{document}

\title{The imaginary part of the high-harmonic cutoff}

\author{Emilio Pisanty\,\orcid{0000-0003-0598-8524}}
 \email{emilio.pisanty@icfo.eu}
 \affiliation{ICFO -- Institut de Ciencies Fotoniques, The Barcelona Institute of Science and Technology, 08860 Castelldefels (Barcelona)}

\author{Marcelo F. Ciappina\,\orcid{0000-0002-1123-6460}}
 \affiliation{ICFO -- Institut de Ciencies Fotoniques, The Barcelona Institute of Science and Technology, 08860 Castelldefels (Barcelona)}

\author{Maciej Lewenstein\,\orcid{0000-0002-0210-7800}\,}
 \affiliation{ICFO -- Institut de Ciencies Fotoniques, The Barcelona Institute of Science and Technology, 08860 Castelldefels (Barcelona)}
 \affiliation{ICREA, Passeig de Lluís Companys, 23, 08010 Barcelona, Spain}

\date{22 July 2020}

\begin{abstract}

High-harmonic generation -- the emission of high-frequency radiation by the ionization and subsequent recombination of an atomic electron driven by a strong laser field -- is widely understood using a quasiclassical trajectory formalism, derived from a saddle-point approximation, where each saddle corresponds to a complex-valued trajectory whose recombination contributes to the harmonic emission.
However, the classification of these saddle points into individual quantum orbits remains a high-friction part of the formalism.
Here we present a scheme to classify these trajectories, based on a natural identification of the (complex) time that corresponds to the harmonic cutoff.
This identification also provides a natural complex value for the cutoff energy, whose imaginary part controls the strength of quantum-path interference between the quantum orbits that meet at the cutoff.
Our construction gives an efficient method to evaluate the location and brightness of the cutoff for a wide class of driver waveforms by solving a single saddle-point equation.
It also allows us to explore the intricate topologies of the Riemann surfaces formed by the quantum orbits induced by nontrivial waveforms.
\\[-2mm]

\noindent
\footnotesize
Accepted Manuscript for
\href{%
  https://doi.org/10.1088/2515-7647/ab8f1e%
  }{%
  \color[rgb]{0,0,0.55}%
  \textit{J. Phys. Photonics} \textbf{2}, 034013  (2020)%
  }, 
available as %
\href{%
  https://arxiv.org/abs/2003.00277%
  }{%
  \color[rgb]{0,0,0.55}%
  arXiv:2003.00277%
  }%
.
\\[-6mm]

\end{abstract}

\maketitle

\setlength{\skip\footins}{18pt plus 5pt}
{\let\thefootnote\relax\footnote{{%
\vspace{2pt}%
\href{https://creativecommons.org/licenses/by/4.0/}{%
\raisebox{-4pt}{\includegraphics[height=1.8em]{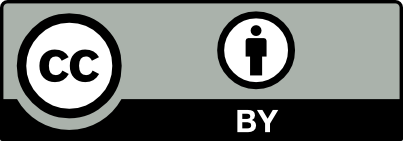}}%
} %
\raisebox{0.5pt}{\textcopyright{}}\,%
The authors, 2020.
Licensed under 
\href{%
  %
  https://creativecommons.org/licenses/by/4.0/%
  }{%
  \color[rgb]{0,0,0.55}%
  CC BY 4.0%
  }.
}}}%
\addtocounter{footnote}{-1}%
High-harmonic generation (HHG) is an extremely nonlinear optical process in which a strong laser field drives the emission of a train of short bursts of high-frequency radiation~\cite{Krausz2009, Corkum2007}, which can cover hundreds of harmonic orders of the driving field, over a broad plateau that terminates at a cutoff.
This emission comes from a three-step process in which the laser ionizes the target atom via tunnel ionization, and then propels the released electron back to the parent ion, where it recombines with the hole it left behind, releasing its kinetic energy as a photon~\cite{Corkum1993, Kulander1993}.

This emission can be modelled using a wide range of approaches, from classical heuristics~\cite{Corkum1993, Kulander1993} to intensive numerical computations~\cite{Scrinzi2014}, but the quantitative models that most closely follow the overall intuition are quasiclassical methods~\cite{Lewenstein1994, Amini2019}, known as the Strong-Field Approximation~(SFA), where the emission amplitude is given by a path-integral sum over discrete emission events.
These are known as quantum orbits~\cite{Salieres2001, Paulus2000, Kopold2002}, i.e., quasiclassical trajectories whose start and end times are complex~\cite{Ivanov2014, Nayak2019}.

The quantum-orbit formalism arises naturally under the approximation that the electron's motion in the continuum is completely controlled by the driving laser, which allows an exact solution in terms of highly oscillatory integrals.
These are then reduced, using a steepest-descent method known as the saddle-point approximation~(SPA), to discrete contributions coming from the saddle points of the integrand~\cite{BruijnAsymptotics, BleisteinIntegrals, GerlachSPAonline}.
These saddle points represent the quasiclassical trajectories, and they typically come in pairs -- most notably the `short' and `long' trajectories~\cite{Lewenstein1995, Zair2008}. 
Each pair of trajectories approaches each other over the harmonic plateau and then performs a Stokes transition at the harmonic cutoff~\cite{Figueira2002, Milosevic2002}, giving way to an exponential drop where only one of the saddles is~used.
In practice, however, classifying the saddle points into these pairs of trajectories is one of the highest-friction points when applying this method~\cite{Chipperfield2007, Hoffmann2011, Das2017}, particularly since the saddles tend to move quickly, and approach each other very closely, at the harmonic cutoff.

\begin{figure}[b]
\begin{tabular}{c}
\subfloat{\label{fig-graph-abs-field} %
\includegraphics[scale=1]{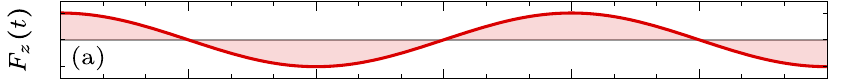} }\\[-3mm]
\subfloat{\label{fig-graph-abs-unsorted} %
\includegraphics[scale=1]{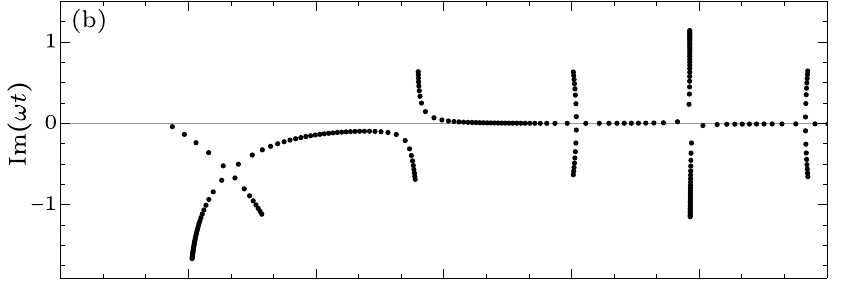} }\\[-3mm]
\subfloat{\label{fig-graph-abs-classified} %
\includegraphics[scale=1]{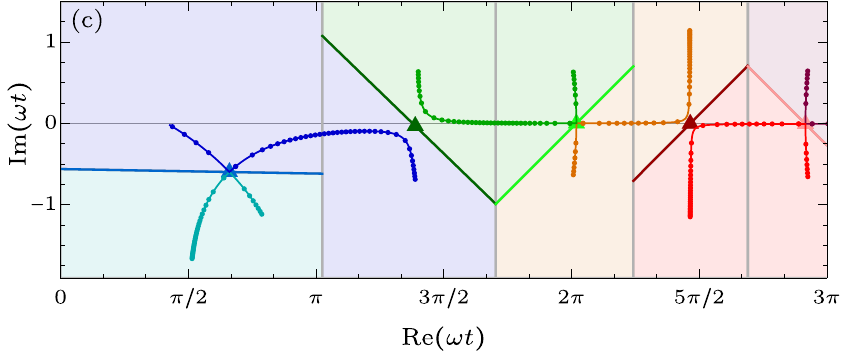} }
\end{tabular}
\caption{
Solving the HHG saddle-point equations returns a discrete set of saddle points as a function of the harmonic photon energy $\Omega$, shown in (b) for the monochromatic field in~(a). 
Our main result is the trajectory classification in~(c), which consists of the harmonic-cutoff points $t_\hc$ (triangles) and the separatrices through them, which allow the cloud of saddle-point solutions in~(b) to be organized into individual trajectories; the different colors correspond to different quantum orbits.
We model neon in a field of wavelength $\SI{800}{nm}$ and intensity $\SI{2e14}{W/cm^2}$.
}
\label{fig-graphical-abstract}
\end{figure}

In this work we construct a quantum-orbit classification scheme based on a natural notion of complex-valued harmonic-cutoff times $t_\hc$. 
These are points -- %
given by zeros of the \textit{second} derivative of the action %
-- which sit between the two saddle points as they approach each other, and which provide an organic separation between the two.
Thus, an unordered set of saddle-point solutions like the one shown in~\reffig{fig-graph-abs-unsorted} can be cleanly organized programmatically into families of trajectories, as shown in~\reffig{fig-graph-abs-classified}, in a flexible and general fashion which is robust to changes in the quantum orbits over broad families of optical fields.

Once these second-order saddle points have been identified, they naturally fill in the role of the quantum orbits corresponding to the high-harmonic cutoff, and they give the photon energy at which it occurs -- a quantity which can often be hard to pin down precisely -- as the real part of the time derivative of the action at the harmonic-cutoff time.
We benchmark this understanding of the harmonic cutoff against the standard `cutoff law', which describes the scaling as $\Omega_\mathrm{max} \sim I_p + 3.17 U_p$ with the ionization potential $I_p$ and the ponderomotive energy $U_p$ of the field (previously derived both classically~\cite{Corkum1993, Kulander1993} and via a systematic expansion in $I_p/U_p$~\cite{Lewenstein1994}).
However, our method extends trivially to drivers with higher harmonic content as well as to tailored polarizations.

Moreover, the information at~$t_\hc$, which can be obtained by solving a single second-order saddle point equation, is sufficient to calculate an accurate evaluation of the harmonic yield at the cutoff, as well as a good qualitative estimate (which we term the Harmonic-Cutoff Approximation) for the shape of the spectrum at the upper plateau.
The efficiency of this approach makes it a good tool when optimizing both the position and the brightness of the cutoff over the high-dimensional parameter spaces available on current optical synthesizers~\cite{Manzoni2015}.

This understanding of the high-harmonic cutoff $\Omega_\hc$ also assigns a natural value to its imaginary part, whose direct impact is to control the closeness of the approach between the pair of saddle points at the cutoff.
Since this closeness regulates, in turn, the distinguishability between the two trajectories, the imaginary part of the harmonic cutoff ultimately controls the strength of quantum-path interference (QPI)~\cite{Zair2008} between the pair of orbits.

In particular, the zeros of the imaginary part of the harmonic cutoff energy $\Im(\Omega_\hc)$ pinpoint the configurations where the saddle points for the two quantum orbits have an exact coalescence at the cutoff, in which case they cannot be distinguished from each other.
For tailored polarizations, as well as other polychromatic fields with one or more nontrivial shape parameters, $\Im(\Omega_\hc)$ will generically oscillate, indicating that the quantum orbits have reconnection events -- where, in essence, the evanescent post-cutoff saddle point will transfer from the `short' quantum orbit to the `long' one.

We showcase this behaviour in bichromatic coun\-ter-ro\-ta\-ting circularly-polarized `bicircular' fields~\cite{Fleischer2014, Kfir2015, Eichmann1995, Milosevic1996, Milosevic2000bicircular, Baykusheva2016, Dorney2017, JimenezGalan2018, Pisanty2014, Pisanty2017, Milovsevic2019} as the relative intensity between the two components changes: the various quantum orbits then recombine with each other, making for a complicated landscape within a unified topology.
This illustrates the fact that the various saddle points are simply different branches of a single, unified Riemann surface, with our harmonic-cutoff times~$t_\hc$ taking on the role of the branch points of this Riemann surface.
More practically, the reconnections make the quantum-orbit classification challenging, but we show that our algorithm can seamlessly handle these changes in the trajectories.

In the more structural terms of catastrophe theory \cite{Poston1978}, the harmonic cutoff is a spectral caustic, in the `fold' class of diffraction catastrophes~\cite{Raz2012}.
In this sense, the complex harmonic-cutoff energy $\Omega_\hc$ is the bifurcation set for the catastrophe -- suitably generalized to the complex variables involved -- and it marks the location of the caustic.
The formal study of caustics has seen increasing interest within attoscience~\cite{Raz2012, Goulielmakis2012, Austin2014, Facciala2016, Facciala2018, Hamilton2017, Birulia2019, Uzan2020, Kelvich2017}, and our approach provides a useful tool for exploring and describing the higher-dimensional bifurcation sets at the heart of the higher-order catastrophes that become available as our control over short pulses of light becomes more finely tuned.

This work is structured as follows.
In Section~\ref{sec-t-hc} we construct the harmonic-cutoff times and examine their structure, first summarizing the standard SFA in Section~\ref{sec-sfa-summary} and constructing a model with a cubic action in Section~\ref{sec-toy-model}, which we explore in depth in Sections~\ref{sec-landscape} and~\ref{sec-saddle-orientation}, where we construct the classification algorithm; in Section~\ref{sec-riemann-surface} we explore the quantum-orbit Riemann surface uncovered by this perspective.
In Section~\ref{sec-yield} we construct the Harmonic-Cutoff Approximation, by explicitly integrating the Airy integral of the cubic action of our model, and we benchmark it against the standard approximations.
In Section~\ref{sec-scaling} we examine how the complex cutoff energy $\Omega_\hc$ scales with the field parameters, and show that it agrees with the known cutoff law and that it extends it to the complex domain.
Finally, in Section~\ref{sec-bicircular}, we explore the branch-cut classification for bicircular fields, showing that our algorithm can handle the quantum-orbit reconnections, as well as how these reconnections give rise to a nontrivial topology for the quantum orbits.

The functionality we describe here has been integrated into the RBSFA package for Mathematica~\cite{RBSFA}, and our specific implementation is available from \citer{FigureMaker}.
Interactive versions of \reffigtd{fig-riemann-surface} and \reffigtd{fig-topology-3D}, as well as 3D-printable models of those surfaces, are available as Supplementary Material~\cite{SupplementaryMaterial}.

\section{The complex harmonic-cutoff times}
\label{sec-t-hc}%
\subsection{The Strong-Field Approximation}
\label{sec-sfa-summary}%
In the SFA, high-harmonic emission is calculated as the expectation value of the dipole operator, under the assumption that there is a single active electron that is either in the atomic potential's ground state $\ket{g}$ or in a laser-driven continuum where the effect of the atomic potential is negligible.
The calculation~\cite{Ivanov2014, Nayak2019, Amini2019, Lewenstein1994} then gives the harmonic dipole in the form
\begin{align}
\vbD(\Omega)
& =
\int_{-\infty}^\infty \!\!\!\!\!\! \d t
\int_{-\infty}^t \!\!\!\!\!\! \d t' \!
\int \! \d\vbp \,
\vbd(\vbp+\vba(t))
\Y(\vbp+\vba(t'))
\nonumber \\ & \qquad \qquad \qquad \qquad \times
e^{-i S_V\!(\vbp,t,t') + i\Omega t}
+\cc
,
\label{sfa-hhg-dipole}
\end{align}
where the three integrals over the times of ionization and recollision, $t'$ and $t$, and the intermediate momentum, $\vbp$, form a reduced Feynman path integral summing over a restricted set of relevant orbits~\cite{Salieres2001}.
This Feynman sum is modulated by ionization and recollision amplitudes given by the transition dipole matrix element $\vbd(\vbk) = \matrixel{\vbk}{\hat{r}}{g}$ and the scalar function $\Y(\vbk) = (I_p + \tfrac12 \vbk^2) \braket{\vbk}{g}$, evaluated at the ionization and recollision velocities: $\vba(t)$ is the vector potential of the laser and $\vbp$ is the canonical momentum, so $\vbv(t)=\vbp+\vba(t)$ is the recollision velocity and $m_e\vbv(t) = \vbv(t)$ is the kinematic momentum. 
(We use atomic units with $m_e=\hbar=1$ throughout, and we consider neon in a linearly-polarized field $\vbf(t) = F_0\ue{z}\cos(\omega t)$ of wavelength $\SI{800}{nm}$ and intensity $\SI{2e14}{W/cm^2}$ unless otherwise stated.)

More importantly, the contribution from each orbit in the Feynman sum in \eqref{sfa-hhg-dipole} is modulated by a phase given by the total action accumulated during propagation,
\begin{subequations}
\begin{align}
S(\vbp,t,t')
& = 
S_V(\vbp,t,t') - \Omega t
\label{sfa-action-total}
\\
\text{for} \ 
S_V(\vbp,t,t') 
& = 
\frac12\int_{t'}^{t} \left(\vbp+\vba(\tau)\right)^2\d\tau + I_p(t-t')
,
\label{sfa-action-volkov}
\end{align}
\label{sfa-action}%
\end{subequations}%
with $I_p$ the ionization potential of the atom, which is normally dominated by the Volkov component $S_V$. For a field with amplitude $F_0$ and frequency $\omega$, the action scales with the ratio $z=U_p/\omega$ of the field's ponderomotive energy $U_p=F_0^2/4\omega^2$ to its frequency.
This is typically large, so the exponential term $e^{-i S}$ changes much faster than the amplitudes in the prefactor, making the integral in \eqref{sfa-hhg-dipole} highly oscillatory.

The highly oscillatory nature of this amplitude then allows us to deal with this integral using the method of steepest descents~\cite{BruijnAsymptotics, BleisteinIntegrals, GerlachSPAonline}, also known as the saddle-point approximation (SPA). 
In this method, we deform the integration contours in~\eqref{sfa-hhg-dipole} into the complex plane, so that they will pass through the saddle points of the exponent~\eqref{sfa-action}.
There the exponent is locally quadratic, so the integral can be reformulated as a gaussian and integrated exactly, under the assumption that the prefactor is slow.
These points can be found through the saddle-point equations
\begin{subequations}
\begin{align}
0 
& = \frac{\partial S}{\partial t} 
  = \frac12\left(\vbp+\vba(t)\right)^2 + I_p - \Omega
,
\label{saddle-point-eqns-t}
\\
0 
& = \frac{\partial S}{\partial t'} 
  = \frac12\left(\vbp+\vba(t')\right)^2 + I_p
,
\label{saddle-point-eqns-tt}
\\
0 
& = \frac{\partial S}{\partial \vbp} 
  = \int_{t'}^{t} \left(\vbp+\vba(\tau)\right)\d\tau
,
\label{saddle-point-eqns-p}
\end{align}
\label{saddle-point-eqns}%
\end{subequations}%
which can be interpreted on physical grounds as encoding the requirements of energy conservation at recollision and ionization (\eqref{saddle-point-eqns-t} and~\eqref{saddle-point-eqns-tt}, resp.) as well as the return of the quasiclassical laser-driven trajectory 
$
\vbalpha(t) = \int_{t'}^t\vbv(\tau)\d\tau
$
to the ion at the time of recollision.%
\footnote{%
We use the term `quasiclassical' here to distinguish this formalism from more general semiclassical approaches, which use Feynman sums over \textit{allowed} classical trajectories and add quantum corrections coming from the gaussian (or other similar) spread of the integral around them.
} %
One key feature of these conditions is that the ionization equation~\eqref{saddle-point-eqns-tt}, in particular, cannot be satisfied with real variables, forcing all of the variables involved to take complex values.

Normally, the return equation~\eqref{saddle-point-eqns-p} is solved separately, since its linearity in $\vbp$ guarantees a unique solution,
\begin{equation}
\vbp_s(t,t')
=
-\frac{1}{t-t'}\int_{t'}^{t} \vba(\tau)\d\tau
,
\end{equation}
for any arbitrary pair~$(t,t')$ of ionization and recollision times.
Once the momentum integral has been performed in this way, the expression for the harmonic dipole takes the form
\begin{align}
\vbD(\Omega)
& =
\int_{-\infty}^\infty \!\!\!\!\! \d t
\int_{-\infty}^t \!\!\!\!\!\! \d t' \:
\vbd(\vbp_s(t,t'){+}\vba(t))
\Y(\vbp_s(t,t'){+}\vba(t'))
\nonumber \\ & \qquad \times
\left(\frac{2\pi}{i(t-t')}\right)^{3/2}
e^{-i S_V\!(t,t') + i\Omega t}
,
\label{sfa-hhg-dipole-p}
\end{align}
where the fractional power of the excursion time $\tau=t-t'$ represents the dispersion of the released wavepacket in position space.%
\footnote{%
If the time integrals are performed numerically, this factor needs to be regularized at $\tau\to 0$, to account for a breakdown of the saddle-point approximation in that limit~\cite{Pisanty2016}. 
If the time integrals are also evaluated via saddle-point methods, however, this is not required, as that limit is not used.%
} %
We notate $S(t,t') = S(\vbp_s(t,t'),t,t')$ where it does not lead to confusion, and we drop the added complex conjugate for simplicity.

The resulting two-dimensional integral,~\eqref{sfa-hhg-dipole-p}, is now in its minimal form; the saddle-point equations for the amended action read
\begin{subequations}
\begin{align}
0 
& = \frac{\partial S}{\partial t} 
  = \frac12\left(\vbp_s(t,t')+\vba(t)\right)^2 + I_p - \Omega
,
\label{saddle-point-eqns-ps-t}
\\
0 
& = \frac{\partial S}{\partial t'} 
  = \frac12\left(\vbp_s(t,t')+\vba(t')\right)^2 + I_p
,
\label{saddle-point-eqns-ps-tt}
\end{align}
\label{saddle-point-eqns-ps}%
\end{subequations}%
and they can only be solved numerically.%
\footnote{%
This is often done by gradient descent on the modulus of the right-hand side~\cite{Ivanov2014, Nayak2019}, but it is also possible to use Newton's method directly, as it readily extends to multiple complex variables~\cite{Chipperfield2007, RBSFA}.
} %
A typical set of solutions for these saddle-point equations is shown in \reffig{fig-graph-abs-unsorted}: the solutions form a discrete set of points, which shift when the harmonic frequency~$\Omega$ changes. 
These discrete points thus trace out individual curves on the complex $t$ and $t'$ planes, which form the individual quantum orbits.

\begin{figure}[t]
\begin{tabular}{c}
\subfloat{\label{fig-traj-approach-field}
\includegraphics[scale=1]{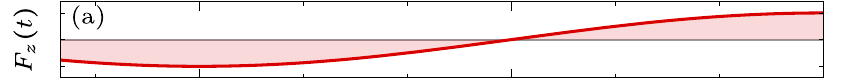} }\\[-3mm]
\subfloat{\label{fig-traj-approach-saddles}
\includegraphics[scale=1]{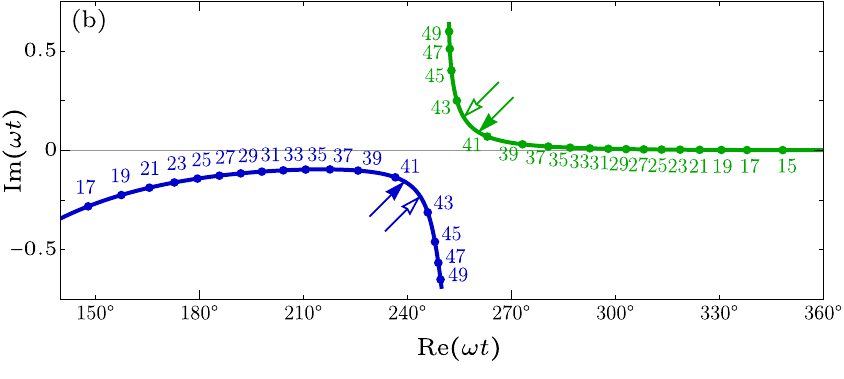} }\\[-2mm]
\subfloat{\label{fig-traj-approach-spectra}
\includegraphics[scale=1]{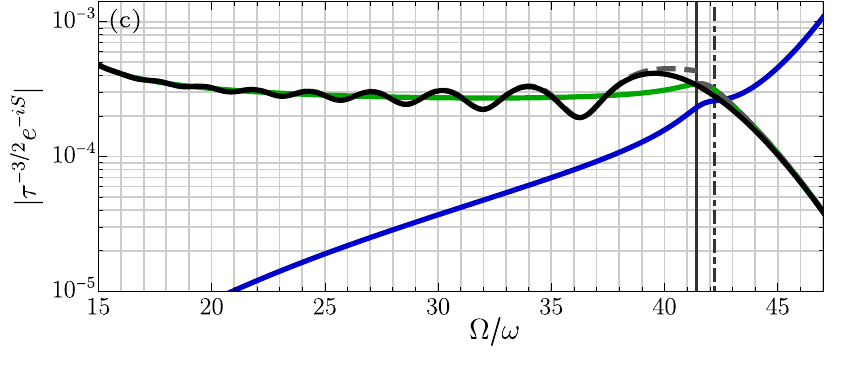} }
\end{tabular}
\caption{
(a,b) Detail of the first pair of quantum orbits from \reffig{fig-graphical-abstract}, labelled by harmonic order, $\Omega/\omega$.
At the avoided cross\-ing of the two saddles, they go through a Stokes and then an anti-Stokes transition (solid and hollow arrow, resp.), at the points where $\Re(S)$ and $\Im(S)$ of the two are equal,~resp.
(c)~Saddle-point approximation to the core elements of the harmonic dipole, $\tau^{-3/2}e^{-iS}$, for the first pair of quantum orbits (green and blue curves).
At the anti-Stokes transition (dot-dashed line), one of the two starts growing exponentially, but it must be discarded at the Stokes transition (solid line), when the required integration-contour topology changes.
This produces a discontinuous change in the total SPA dipole (dashed line), which can be fixed by using the Uniform Approximation (solid line).
The oscillations over the plateau are quantum path interference between the two trajectories.
}
\label{fig-traj-approach}
\end{figure}

Once the saddle points have been found, the SPA gives the harmonic dipole as
\begin{align}
\vbD(\Omega)
& =
\sum_s
\frac{2\pi}{\sqrt{-\det[S''(t_s,t'_s)]}}
\left(\frac{2\pi}{i(t_s-t'_s)}\right)^{3/2}
\nonumber \\ & \qquad \qquad \times
\label{spa-dipole}
\vbd(\vbp_s{+}\vba(t_s))
\Y(\vbp_s{+}\vba(t'_s))
\,
\\ & \qquad \qquad \times \nonumber
e^{-i S_V\!(t_s,t'_s) + i\Omega t_s}
,
\end{align}
with $\det[S''] = \partial_{tt}^2S \, \partial_{t't'}^2S - (\partial_{t t'}^2 S)^2$ the Hessian of the action, thus giving the harmonic yield as a coherent sum of discrete contributions coming from each of the quantum orbits.
\reffig{fig-traj-approach-spectra} shows a typical example of how the individual contributions (blue and green curves) get combined into a total harmonic yield, with quantum-path interference beatings in the plateau as the two contributions go in and out of phase~\cite{Lewenstein1995, Zair2008}.

Within the SPA expression~\eqref{spa-dipole}, the key component is the summation: this runs over all the saddle points of the action which are compatible with a suitable steepest-descent integration contour, a property that can be nontrivial to determine.
\reffig{fig-traj-approach-saddles} shows a typical configuration, with the short- and long-trajectory saddle points approaching each other over the harmonic plateau, close to the real axis, performing an `avoided crossing' at the harmonic cutoff, and then advancing into imaginary time. 

At this avoided crossing, the saddle-point pair experiences two important transitions, known as the Stokes and anti-Stokes lines~\cite{Figueira2002, Milosevic2002, Chipperfield2007}.
At the anti-Stokes transition, which is defined as the point where $\Im(S)$ for the two orbits is equal, the short-trajectory contribution begins to grow exponentially, and it must be discarded from the summation in order to keep reasonably physical results.

This elimination is enforced by the Stokes transition, which typically happens earlier, defined as the point where $\Re(S)$ for the two orbits is equal and then changes order.
This is an important change, as the steepest-descent method requires integration contours to follow the contour lines of $\Re(S)$: thus, the change in the ordering of $\Re(S(t_s,t_s'))$ for the long- and short-trajectory saddle points means that, after the Stokes transition, the short-trajectory saddle point (in this example) can no longer form part of a suitable integration contour, and it needs to be discarded from the summation.

This means, however, that the SPA harmonic yield at the cutoff is a discontinuous function of $\Omega$, coming from the discrete jump at the points where one of the trajectories is eliminated, and this discontinuity is clearly incompatible with the initial, obviously continuous, expressions for $\vbD(\Omega)$ in~\eqref{sfa-hhg-dipole} and~\eqref{sfa-hhg-dipole-p}.
This apparent paradox is resolved by noting that the SPA is not valid when saddles are close together, as the quadratic approximation to the exponent fails.
Instead, one must use a cubic form for the exponent, which can be integrated in terms of Airy or fractional-order Bessel functions.
This gives a continuous spectrum from the saddle-point pair, known as the Uniform Approximation~\cite{Figueira2002, Milosevic2002} (UA), which is shown in~\reffig{fig-traj-approach-spectra}.

From a more general viewpoint, it is important to stress that applying these considerations is only possible once the various saddle points have been classified into continuously-connected quantum orbits, as without that step it is impossible to even define the objects that will be included in the summation or discarded from~it.%
\footnote{%
Similarly, keeping good track of the saddle points in continuous quantum orbits is also essential to ensure that the Hessian square root in~\eqref{spa-dipole}, $\sqrt{-\det[S''(t_s,t'_s)]}$, does not cross any branch cuts.
} %
In practice, this is a high-friction point in the calculation, requiring expensively tight grid spacings in $\Omega$ to accurately resolve the avoided crossings, or the additional design of an adaptive grid to increase the energy resolution there~\cite{Chipperfield2007, Hoffmann2011, Das2017}. 
Moreover, this problem is compounded in more complex polychromatic fields, where the saddle-point structure changes depending on the details of the laser pulse.
It is the goal of this manuscript to provide a simple and effective method to classify these quantum orbits, separating the points in~\reffig{fig-traj-approach-saddles} along the diagonal into the two curves shown.

\subsection{A model for the saddle points at the cutoff}
\label{sec-toy-model}
In order to build this method, we first consider a simple model for the saddle points at and near the harmonic cutoff, in order to isolate the key features of the problem.
In essence, \reffig{fig-traj-approach-saddles} shows us two saddle points approaching each other and then receding, so we consider the simplest model action with only two saddles, of the polynomial form
\begin{equation}
\frac{\d S}{\d t} = A(t-t_{s,1})(t-t_{s,2})
.
\label{model-action-1}
\end{equation}
(For simplicity, we restrict our attention for now to actions with only one variable, $t$, which corresponds to solving~\eqref{saddle-point-eqns-ps-tt} first for $t_s'=t_s'(t)$, giving $S(t)=S(t,t_s'(t))$ and then examining~\eqref{saddle-point-eqns-ps-t}. We shall lift this restriction later.)
Moreover, we use the clear symmetry evident in \reffig{fig-traj-approach-saddles}, with both saddles (approximately) symmetrically placed about the center of the plot, and enforce the condition $t_{s,1}=-t_{s,2}$ in our model, so that its action obeys
\begin{align}
\frac{\d S}{\d t} 
& = A(t-t_s)(t+t_s)
\nonumber \\
& = A(t^2-t_s^2)
,
\label{model-action-2}
\end{align}
which can be integrated to find
\begin{equation}
S(t)
=
\frac13 A \, t^3 - At_s^2 \, t + C
\label{model-action-3}
\end{equation}
as a cubic polynomial, where we set $C=0$ for simplicity.

Turning now to the oscillatory integral for this action, we can define it in the form
\begin{align}
\int e^{-iS(t)} \d t
=
\int e^{-\frac i3 A t^3 +i A t_s^2 \, t} \, \d t
\label{model-action-4}
\end{align}
and then compare the linear term, $e^{+i A t_s^2 \, t}$, with the Fourier kernel $e^{+i\Omega t}$ from \eqref{sfa-hhg-dipole}, which has the same structure.
Thus, in order to turn \eqref{model-action-4} into a form that is clearly analogous to \eqref{sfa-hhg-dipole}, we separate $A t_s^2 = \Omega-(\Omega_c+i\eta)$ into a variable and a constant part, and thus define
\begin{align}
D(\Omega)
=
\int e^{-\frac i3 A t^3 +i (\Omega-\Omega_c-i\eta) t} \, \d t
.
\label{model-action-5}
\end{align}
Here the constant part $\Omega_\hc = \Omega_c+i\eta$ has a nonzero imaginary part $+i\eta$, as we do not have any guarantees that $At_s^2$ is real, but its real part $\Omega_c$ can be set to zero if desired, since it simply acts as an offset for the variable~$\Omega$.
Here the functional dependence 
\begin{equation}
A t_s^2 = \Omega-(\Omega_c+i\eta) = \Omega - \Omega_\hc
\label{Omega-hc-definition}
\end{equation}
on $\Omega$ is an additional postulate, justified only by analogy with the Fourier kernel of the full integral which our model attempts to mimic, but, as we shall see shortly, the `quantum orbits' $t_s(\Omega)$ that result from this identification form a good model for the avoided-crossing behaviour in~\reffig{fig-traj-approach-saddles}.

In addition, if we now separate $S(t) = S_V(t) -\Omega t$ as we did in~\eqref{sfa-action} above, the saddle-point equation~\eqref{model-action-2} can now be rephrased as the requirement that
\begin{subequations}%
\begin{align}
\frac{\d S_V}{\d t} 
= At^2 + (\Omega_c + i\eta)
= \Omega
,
\label{model-action-6}
\end{align}
with $\Omega$ running over the real axis, in direct analogy to~\eqref{saddle-point-eqns-t} and~\eqref{saddle-point-eqns-ps-t}.
This is our first key insight: the curves traced out by the solutions of the saddle-point equation~\eqref{model-action-2} can also be described as the contour line 
\begin{equation}
\Im\mathopen{}\left[\frac{\d S_V}{\d t} \right]\mathclose{} = 0
\label{model-action-contour}
\end{equation}
\end{subequations}%
of the derivative of the model Volkov action~$S_V$ (and also of the full action $S$, since they differ by a real number).

\begin{figure}[t]
\begin{tabular}{c}
\includegraphics[scale=1]{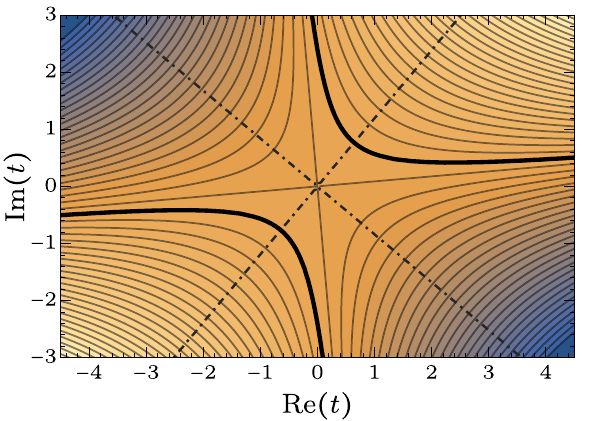} 
\end{tabular}
\caption{
Contour map of $\Im\mathopen{}\left[\frac{\d S_V}{\d t} \right]\mathclose{}$ for the model action in~\eqref{model-action-3}, with blue (yellow) indicating negative (positive) values, and with the $\Im\mathopen{}\left[\frac{\d S_V}{\d t} \right]\mathclose{} = 0$ contour highlighted in black.
The gray dot-dashed lines are the contours of $\Re\mathopen{}\left[\frac{\d S_V}{\d t} \right]\mathclose{}$ passing through the central saddle point.
We set $\eta=-1$ and $A=e^{\SI{10}{\degree}i}$.
}
\label{fig-model-contour-map}
\end{figure}

This insight then lights the way further: our principal task is to look for objects between the two curves in \reffig{fig-traj-approach-saddles} that will help us separate them, and the contour map of $\Im\mathopen{}\left[\frac{\d S_V}{\d t} \right]\mathclose{}$ is clearly the place to look.
We show this contour map in \reffig{fig-model-contour-map}, with the zero contour of~\eqref{model-action-contour} highlighted as a thick black curve, clearly showing the avoided-crossing structure of \reffig{fig-traj-approach-saddles} that we want to model.
More importantly, this plot shows us our second key insight: there is indeed a nontrivial object separating the two quantum orbits, in the form of a saddle point in this contour map.
This is the central object we are after: the harmonic-cutoff time $t_\hc$ for this model. 
Like all saddle points, this can be found as a zero of the derivative of the contour map's original function, or in other words,~via
\begin{equation}
\frac{\d^2 S_V}{\d t^2} (t_\hc) = 0
.
\label{model-action-thc-equation}
\end{equation}
In the particular case of our model action~\eqref{model-action-6}, for which $\frac{\d^2 S_V}{\d t^2} =2A \, t$, this saddle point lies at the origin, due to the explicit choice of center of symmetry made by setting $t_{s,1}=-t_{s,2}$ above. 
In the general case,~\eqref{model-action-thc-equation} will find the centerpoint between $\Im\mathopen{}\left[\frac{\d S_V}{\d t} \right]\mathclose{}$ contours whenever they approach each other.

The appearance of a saddle point at the midpoint between contour lines in close proximity is a generic feature of contour landscapes.
Similar structures have been explored previously~\cite{Pisanty2016slalom, Keil2016}, in the context of tunnel ionization.

\subsection{The landscape at the harmonic-cutoff point}
\label{sec-landscape}
Having found the saddle point, we can now use it directly, since we can fully reconstruct the model action $S_V(t)$ using only its behaviour at the center, by taking derivatives:
\begin{subequations}%
\begin{align}
\frac{\d S_V}{\d t} (t_\hc) & = \Omega_\hc = \Omega_c + i \eta,
\label{action-derivatives-thc-1}
\\
\frac{\d^3 S_V}{\d t^3} (t_\hc) & = 2A,
\label{action-derivatives-thc-3}
\end{align}
\label{action-derivatives-thc}%
\end{subequations}%
with the second derivative vanishing by construction.

\begin{figure}[t]
\begin{tabular}{c}
\includegraphics[scale=1]{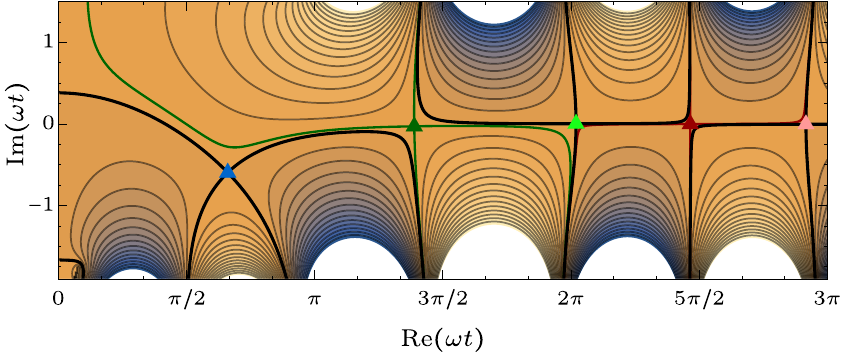} 
\end{tabular}
\caption{
Contour map of $\Im\mathopen{}\left[\frac{\partial S_V}{\partial t}(t,t_s'(t))\right]$ over the complex recollision-time plane for the full Volkov action from~\eqref{sfa-action-volkov}.
The saddle-point trajectories of \reffig{fig-graphical-abstract} appear clearly as the zero contour (thick lines), i.e.\ as the complex times where $\frac{\partial S_V}{\partial t}(t,t_s'(t))$ is real-valued.
The harmonic-cutoff points of \reffig{fig-graph-abs-classified} (triangles) are, correspondingly, the saddle points of this derivative; the coloured lines show the contours at these saddle points.
}
\label{fig-dSdt}
\end{figure}

We begin by focusing on the first derivative, which gives the linear term in the action, whose coefficient is directly connected to the harmonic frequency via $A t_s^2 = \Omega-(\Omega_c+i\eta)$ as set above.
In particular, this term directly encodes the saddle-point solutions, and we can recover these explicitly by inverting that relationship:
\begin{equation}
t_s = \pm \sqrt{\frac{\Omega-(\Omega_c+i\eta)}{A}}
.
\label{ts-explicit-sqrt}
\end{equation}
Here the square in $A t_s^2$ implies that $t_s$ appears as an explicit square root, with the sign ambiguity giving us the two saddle-point solutions of the problem.

This is also a key structural insight, which has largely remained unobserved in the literature: the various solutions of saddle-point equations are simply different branches of a unified Riemann surface, examined at the real axis of the target space. 
This observation is made explicit in the branch choice given by the $\pm$ sign in~\eqref{ts-explicit-sqrt}, but the same is true in the full problem: 
if we approach the coupled equations in~\eqref{saddle-point-eqns-ps} by solving~\eqref{saddle-point-eqns-ps-tt} first for $t_s'=t_s'(t)$, then~\eqref{saddle-point-eqns-ps-t} can be expressed in the form
\begin{equation}
\frac{\partial S_V}{\partial t}(t,t_s'(t)) 
= \frac12\left(\vbp_s(t,t_s'(t))+\vba(t)\right)^2 + I_p 
= \Omega
,
\label{dSdt-many-to-one}
\end{equation}
where it explicitly involves the inverse of the many-to-one, complex-valued function $\frac{\partial S_V}{\partial t}(t,t_s'(t))$, and this is precisely the type of problem encoded by Riemann surfaces.
(We explore this perspective in detail in Section~\ref{sec-riemann-surface} below.)

Moreover, as far as the full HHG problem is concerned, our insight in~\eqref{model-action-contour} above that the saddle-point trajectories are simply contour lines of a derivative of the action remains unchanged, with the obvious alterations to
\begin{equation}
\Im\mathopen{}\left[\frac{\partial S_V}{\partial t}(t,t_s'(t)) \right]\mathclose{} = 0
.
\end{equation}
We exemplify this directly in \reffig{fig-dSdt}, showing the contour map of $\Im\mathopen{}\left[\frac{\partial S_V}{\partial t}(t,t_s'(t)) \right]\mathclose{}$ for the full HHG problem as in \reffig{fig-graphical-abstract}, and there the zero contour lines (highlighted in black) are indeed the curves followed by the typical quantum orbits shown in~\reffig{fig-graph-abs-classified}.
Similarly, the harmonic-cutoff times $t_\hc$ used there are found as the saddle points of the $\frac{\partial S_V}{\partial t}(t,t_s'(t)){}$ landscape in \reffig{fig-dSdt}.

That said, the quantum-orbit curves of \reffig{fig-dSdt} are not simply unstructured curves, as they have an explicit parametrization in terms of the harmonic frequency $\Omega$, but the same is true for the contours of $\Im\mathopen{}\left[\frac{\d S_V}{\d t} \right]\mathclose{}$, which can also be seen as parametrized by the real part of the function.
This brings physical content to the first part of the derivatives we found in~\eqref{action-derivatives-thc}, where the real part of $\frac{\d S_V}{\d t}(t_\hc)$ is simply the frequency offset $\Omega$.
This offset is carried from the saddle point $t_\hc$ to the quantum-orbit lines by means of the contour lines of $\Re\mathopen{}\left[\frac{\d S_V}{\d t} \right]\mathclose{}$.
These are shown dot-dashed in \reffig{fig-dSdt}, and they clearly intersect the quantum-orbit tracks at the point where they are closest to each other, and thus at the transition itself.
This is then what allows us to identify
\begin{subequations}%
\begin{equation}
\Omega_c = \Re\mathopen{}\left[\frac{\d S_V}{\d t}(t_\hc) \right]\mathclose{}
\label{Re-Omega-hc-setting}
\end{equation}
as the frequency of the harmonic cutoff.

Having made this leap, we must now confront the fact that $\frac{\d S_V}{\d t} (t_\hc)$ as given in~\eqref{action-derivatives-thc-1} also has a nonzero imaginary part,
\begin{equation}
\eta = \Im\mathopen{}\left[\frac{\d S_V}{\d t}(t_\hc) \right]\mathclose{},
\label{Im-Omega-hc-setting}
\end{equation}
\label{Omega-hc-setting}%
\end{subequations}%
which similarly needs to be addressed. 
This imaginary part has already played a role, and it is explicitly shown in \reffig{fig-model-contour-map} as the height of the $\frac{\d S_V}{\d t}$ saddle with respect to the zero contour where the quantum orbits lie.
As such, the imaginary part $\eta$ of the harmonic cutoff, defined by~\eqref{action-derivatives-thc-1}, controls how closely the two quantum orbits approach each other at the cutoff, and thus how distinguishable they are. 
As we shall see in Section~\ref{sec-airy-model-qpi} below, this distinguishability further controls the strength of quantum-path interference between them.

Moreover, the sign of $\eta$ controls the direction in which the quantum orbits turn when they approach each other, and thus which of the two post-cutoff evanescent solutions, at positive and negative imaginary time, corresponds to the `short' and `long' trajectory to the left and right of $t_\hc$. 
Generally, this imaginary part of $\frac{\d S_V}{\d t}$ is fixed by the problem and it can be either positive or negative, but, as we shall see in Section~\ref{sec-bicircular}, sign changes in $\eta$ are generic behaviour when the driving pulse shape changes depending on one or more parameters, which implies reconnections and changes in topology for the curves traced out by the quantum orbits.
This further emphasizes the fact that the different saddle points are essentially one and the same object, corresponding only to different branches of one unified Riemann surface.

However, this point can be observed more cleanly, by simply allowing the harmonic frequency $\Omega$ to acquire complex values, i.e., by considering the analytical continuation of $D(\Omega)$. 
Doing this amounts to directly affecting the value of $\eta$ in~\eqref{model-action-6}, and if $\Omega$ is taken at the complex harmonic cutoff itself, $\Omega_c+i\eta$, then the constant term in the saddle-point equation vanishes, leaving a double zero of the form
\begin{equation}
At^2 = 0
\end{equation}
at $t=t_\hc$.
In other words, the harmonic-cutoff times~$t_\hc$ are the locations where the saddle points for the two quantum orbits fully coalesce into a single point.

\subsection{The orientation of the harmonic-cutoff saddle and quantum-orbit classification}
\label{sec-saddle-orientation}
Having examined the first derivative of $S_V$ at the harmonic-cutoff time, we now turn to the role of the third derivative, given in~\eqref{action-derivatives-thc-3}.
This term gives the coefficient of the quadratic term in $\frac{\d S_V}{\d t}$ and, as such, it controls the orientation of the saddle in $\frac{\d S_V}{\d t}$ at $t_\hc$, as shown in \reffig{fig-model-contour-map}. 
(Indeed, we chose a nonzero phase for $A=e^{\SI{10}{\degree}i}$ for that configuration to emphasize that this saddle need not be neatly oriented, as observed in \reffig{fig-dSdt}.)
Retrieving this orientation is essential in order to use our harmonic-cutoff times $t_\hc$ to classify the saddle points into quantum orbits: in order to turn the clouds of points in \reffig{fig-graph-abs-unsorted} into the ordered curves of \reffig{fig-graph-abs-classified}, we also require the direction of the separatrix that goes through the missed approach.

To obtain this direction, we look for the vector that goes from $t_\hc$ to the closest point on the $\Im\mathopen{}\left[\frac{\d S_V}{\d t} \right]\mathclose{} = 0$ contour that the quantum orbits follow.
As we noted above, this point occurs at $\Omega=\Omega_c$, which means that we simply need to solve the saddle-point equation~\eqref{model-action-6} at that frequency,~i.e.,
\begin{align}
At_\sep^2 + i\eta = 0
\ \implies \ 
t_\sep = \sqrt{-i\eta/A}
,
\end{align}
or, in terms of the explicit derivatives of the action and with an explicit relation to the saddle center $t_\hc$,
\begin{align}
\delta t_\sep
=
t_\sep - t_\hc
= 
\sqrt{-2i 
  \frac{
    \Im\mathopen{}\left[
      \displaystyle
      \frac{\d S_V}{\d t}(t_\hc) 
      \right]\mathclose{}
    }{
    \displaystyle
    \frac{\d^3 S_V}{\d t^3} (t_\hc)
    }
  }
.
\label{tsep-definition}
\end{align}
We can then write the explicit condition for the separatrix by treating $\delta t_\sep$ and $t_s-t_\hc$ as vectors and asking for the sign their inner product. 
Thus, the criterion that separates the two quantum orbits on either side of a given harmonic-cutoff time $t_\hc$ is the sign comparison
\begin{equation}
\Re\mathopen{}\Big[
  (t_s - t_\hc)^* \, \delta t_\sep
  \Big]\mathclose{}
\lessgtr 0
,
\label{separatrix-condition}
\end{equation}
with $t_\hc$ defined as in~\eqref{model-action-thc-equation} and $\delta t_\sep$ defined as in~\eqref{tsep-definition}.

\begin{figure}[t]
\begin{tabular}{c}
\includegraphics[scale=1]{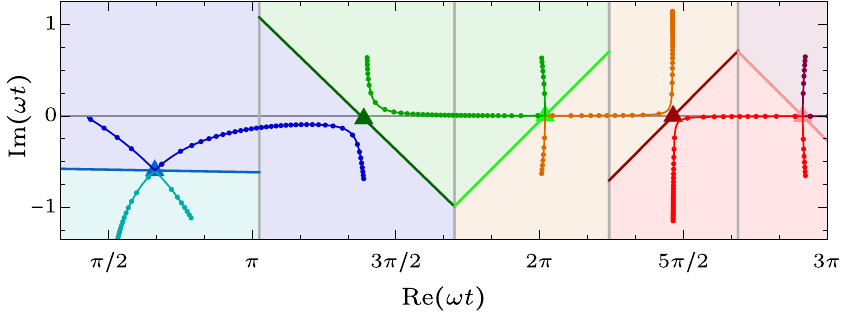} 
\end{tabular}
\caption{
Results of our saddle-point classification procedure, as shown previously in \reffig{fig-graph-abs-classified}.
We find harmonic-cutoff times $t_\hc$ (triangles) as solutions of \eqref{model-action-thc-equation}, with the separatrix through each $t_\hc$ given by the sign comparison in \eqref{separatrix-condition} with $\delta t_\sep$ defined in \eqref{tsep-definition}; we apply this criterion on vertical strips obtained by taking the midpoints between the real parts of the $t_\hc$.
This then defines clear regions occupied by each of the quantum orbits, which can be unambiguously labelled.
}
\label{fig-saddle-classification}
\end{figure}

In a practical calculation, there will typically be multiple harmonic-cutoff times, alternating between near-threshold harmonic frequencies, where the quantum orbits first appear, and high-frequency cutoffs.
To complete our classification scheme -- shown as the background coloured zones in \reffig{fig-graph-abs-classified}, which we repeat in some additional detail in \reffig{fig-saddle-classification} -- we separate the complex time plane into strips using the mid-points between the real parts of the successive harmonic-cutoff times, and then divide each strip in two using the criterion in~\eqref{separatrix-condition}.

Here it is important to note that the definition in \eqref{tsep-definition} contains a sign ambiguity coming from the choice of sign for the square root.
In practice, the principal branch of the square root tends to work well for most cases, but the branch choice there does occasionally require dedicated attention, particularly for the first low-energy harmonic cutoff at the start of the series.

The results of our classification procedure are shown in \reffig{fig-saddle-classification}: the separatrices obtained from the sign-comparison criterion \eqref{separatrix-condition} break up the complex plane into trapezoids that contain one and only one quantum orbit each, and they can thus be used to classify the saddle-point solutions into well-defined families in a uniform and robust fashion.

\begin{tdfigure*}[t]
\newlength{\figureOdownwidth}
\setlength{\figureOdownwidth}{0.49\columnwidth}
\begin{tabular}{c}
\subfloat{\label{fig-riemann-surface-re}
  \href{https://imaginary-harmonic-cutoff.github.io/\#figure-1ab}{
  \includegraphics[scale=0.5]{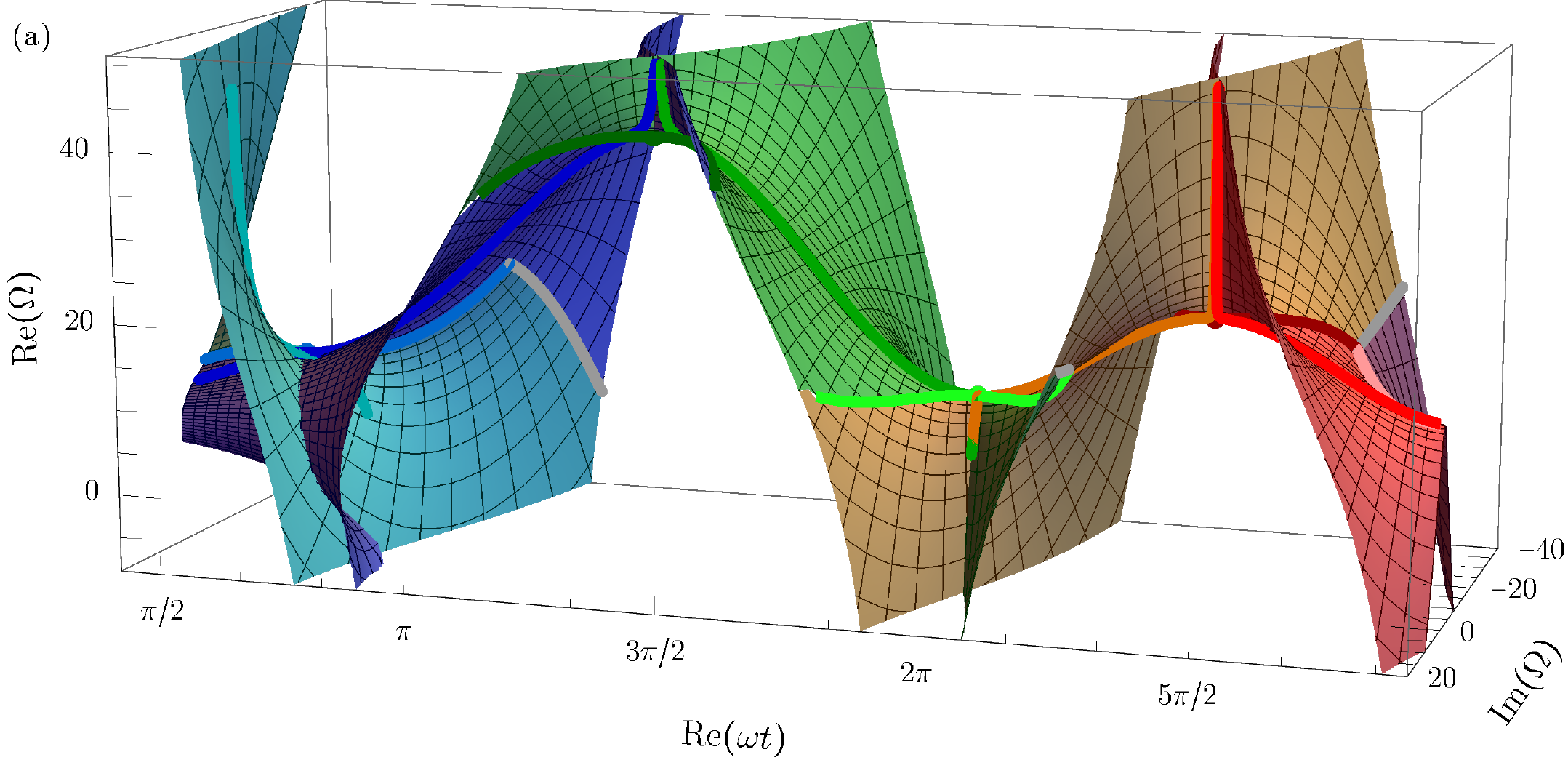} }}
\subfloat{\label{fig-riemann-surface-re-detail}
  \href{https://imaginary-harmonic-cutoff.github.io/\#figure-1ab}{
  \includegraphics[scale=0.5]{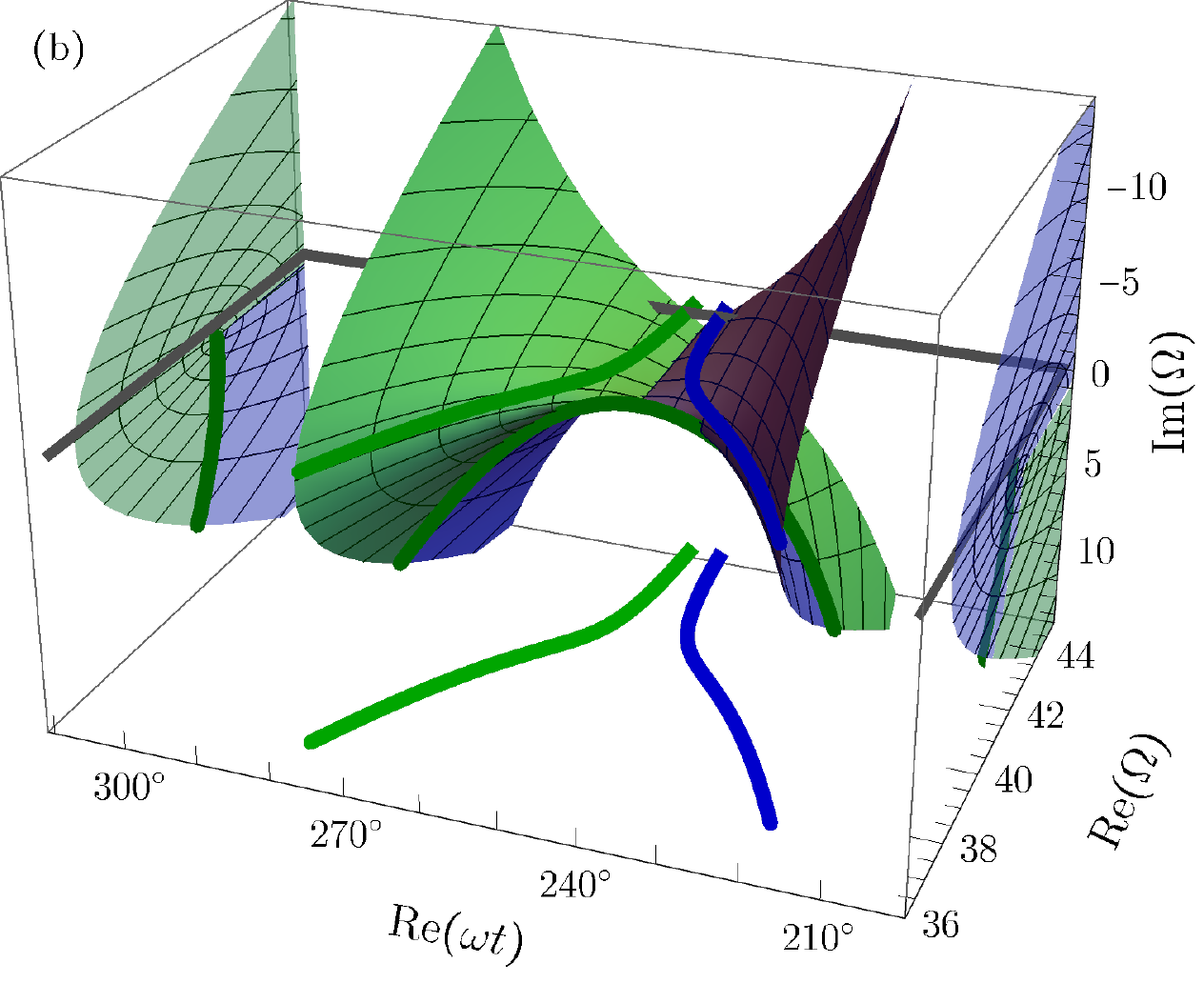} }}
\\[4mm]
\subfloat{\label{fig-riemann-surface-im-1}
  \href{https://imaginary-harmonic-cutoff.github.io/\#figure-1c}{
  \includegraphics[scale=0.5]{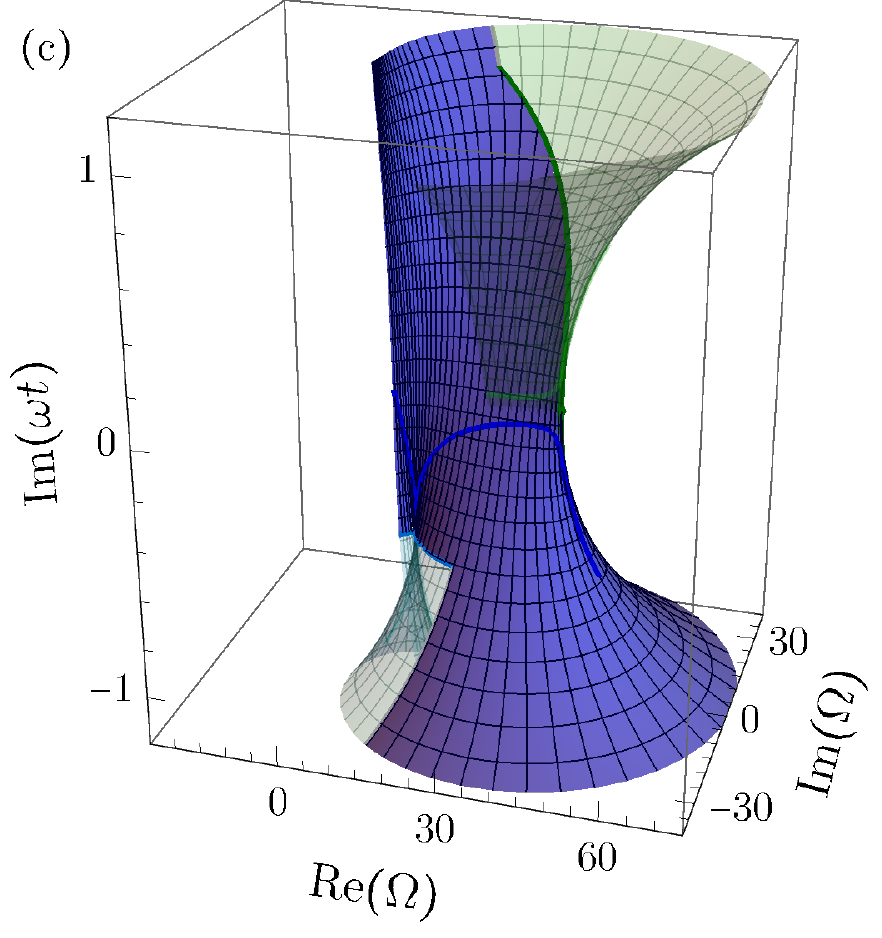} }}
  \hspace{-4.5mm}
\subfloat{\label{fig-riemann-surface-im-2}
  \href{https://imaginary-harmonic-cutoff.github.io/\#figure-1d}{
  \includegraphics[scale=0.5]{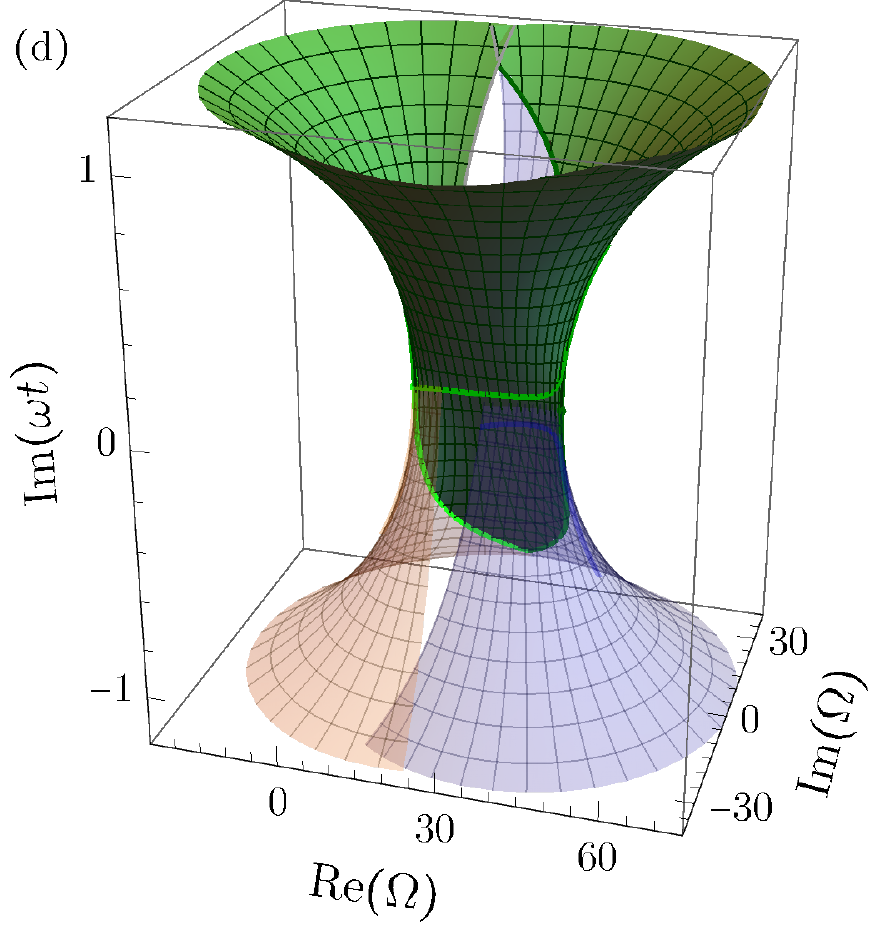} }}
  \hspace{-4.5mm}
\subfloat{\label{fig-riemann-surface-im-3}
  \href{https://imaginary-harmonic-cutoff.github.io/\#figure-1e}{
  \includegraphics[scale=0.5]{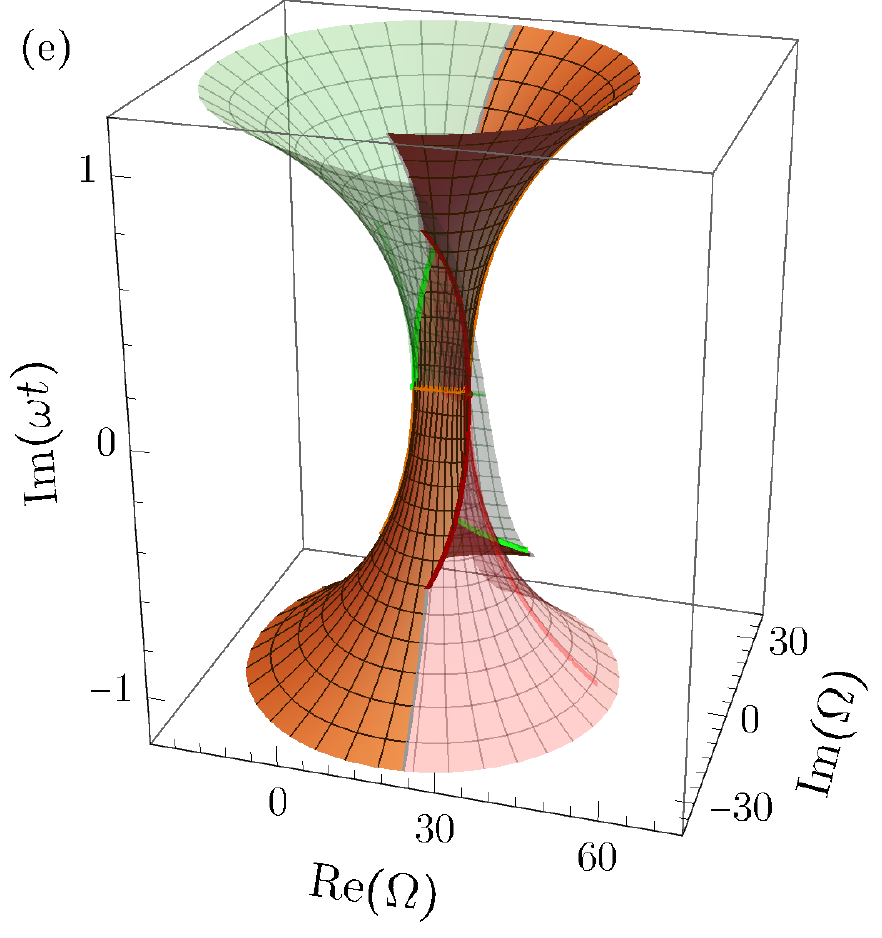} }}
  \hspace{-4.5mm}
\subfloat{\label{fig-riemann-surface-im-4}
  \href{https://imaginary-harmonic-cutoff.github.io/\#figure-1f}{
  \includegraphics[scale=0.5]{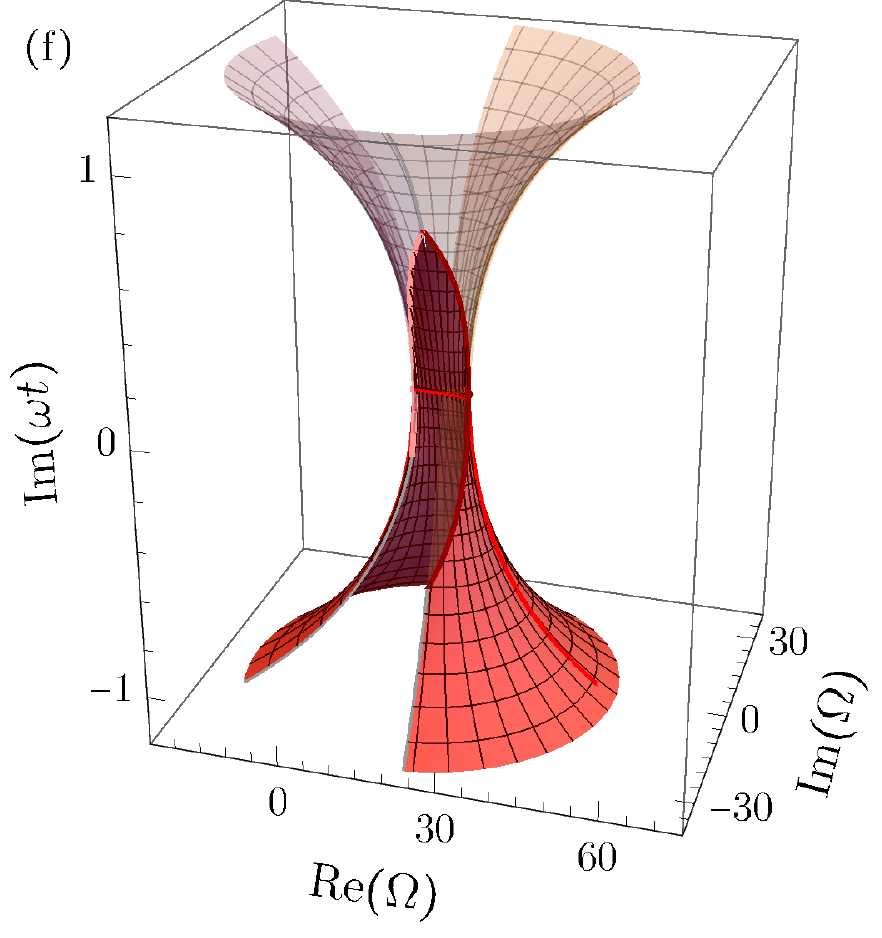} }}
\\[1mm]
\end{tabular}
\caption{
Topology of the Riemann surface of the quantum orbits.
This is the manifold $\mathcal S = \{(t,\Omega)\in\mathbb{C}^2: \Omega = \frac{\d S_V}{\d t}(t)\}$, which has complex dimension~1 and is embedded in the space $\mathbb{C}^2$, with real dimension~4, so we can only plot it by projecting out one component:
we show its projections to $\Re(\omega t)$ in (a,\,b) and to $\Im(\omega t)$ in (c-f).
The projection to $\Re(\omega t)$ shows the topology as a single sheet which wraps around itself, covering the $\Omega\in\mathbb C$ plane multiple times when the $\Re(\omega t)$ coordinate is projected out, so that $\frac{\d S_V}{\d t}$ has a multi-valued inverse.
To separate this multi-valued inverse into valid single-valued functions, the sheet needs to be cut into separate branches: this is the role of the separation into coloured regions from the classification scheme in \reffig{fig-saddle-classification}, which carry over to the multiple coloured sections in (a).
The sheets connect to each other via square-root-type branch cuts, and we show a detail of the first such connection in (b).
The projection onto $\Im(\omega t)$ forms a self-intersecting surface which loops around itself multiple times, so we plot each of the branches separately (with half of each of its neighbours shown half-transparently) in (c-f).
Interactive versions and 3D-printable models of these plots are available at \href{https://imaginary-harmonic-cutoff.github.io}{imaginary-harmonic-cutoff.github.io}~\cite{SupplementaryMaterial}.
%
}
\label{fig-riemann-surface}
\end{tdfigure*}

\subsection{The quantum-orbit Riemann surface}
\label{sec-riemann-surface}
We now return to an observation we made earlier: the saddle-point equations for HHG represent the inverse of a many-to-one analytical function, $\frac{\partial S_V}{\partial t}$ as given in \eqref{dSdt-many-to-one}, evaluated on the real axis in $\Omega$, and this is precisely the definition of a Riemann surface. 
As such, it is pertinent to study this Riemann surface as a whole, since its topology and geometry are the key factors that govern the quantum-orbit classification.

The Riemann surface here is the set
\begin{equation}
\mathcal{S} 
= 
\left\{
  (t,\Omega)\in\mathbb C^2 : 
  \Omega = \frac{\d S_V}{\d t}(t) 
  \right\},
\end{equation}
and it forms a manifold of complex dimension 1 (i.e., a manifold where each point has a neighbourhood homeomorphic with the complex unit disk) embedded in a space of complex dimension 2 (and therefore of real dimension~4). 
This space is too big to be visualized directly, so we approach it by projecting down to the real and imaginary parts of $t$, i.e., by projecting the surface to the spaces $(\Re(\Omega), \Im(\Omega), \Re(\omega t))$ and $(\Re(\Omega), \Im(\Omega), \Im(\omega t))$, which we show in \reffigtd{fig-riemann-surface}.

The topology is most clearly displayed in the $\Re(\omega t)$ projection, shown in \reffigtdsub{fig-riemann-surface}{fig-riemann-surface-re}: the Riemann surface consists of a series of connected sheets which connect sequentially to each other, one by one, as $\Re(\omega t)$ increases.
In essence, the topology and the overall geometry here are those of the Riemann surface of $\sin(z)$, with the sheets connected pairwise by square-root branch points (locally homeomorphic to the quadratic model discussed in Section~\ref{sec-toy-model}) as shown in detail in \reffigtdsub{fig-riemann-surface}{fig-riemann-surface-re-detail}.

This Riemann surface carries an image of the $\omega t$ complex plane (so, in particular, the mesh on the surface represents a square grid on that plane), but it forms a multiple cover of the image plane $\Omega\in\mathbb C$. 
To get the inverse, then, the surface must be split into separate branches, and this is precisely the role of the coloured regions shown in \reffig{fig-saddle-classification} as produced by our saddle-point classification algorithm: this colouring is retained in the Riemann surface as displayed in \reffigtd{fig-riemann-surface}, and each individual region, when projected down to the $\Omega$ plane, forms a single full cover of the complex plane.%
\footnote{%
This single-cover property is approximate, as there is still a small amount of double-cover overlap in regions next to the separatrix when away from the branch points at the $t_\hc$. This can be fixed if necessary, but it is not central to our argument here.
} %
In other words, each of the coloured regions in \reffig{fig-saddle-classification} forms the image of a single-valued branch of the inverse of $\frac{\d S_V}{\d t}(t)$.

The imaginary-part projection onto $\Im(\omega t)$ is slightly harder to represent and visualize, as the corresponding surface folds back on itself, with multiple self-inter\-sec\-tions. 
However, the separation of this surface into individual branches also solves this problem: when plotting one branch at a time, as we show in \reffigtdsub{fig-riemann-surface}{fig-riemann-surface-im-1}-\ref{fig-riemann-surface-im-4}, the self-intersections disappear -- as they must if the surface represents a single-valued function -- and they are only present as intersections between different branches.

Physically, the most important part of this Riemann surface is its intersection with the $\Omega \in \mathbb R^+$ positive real axis in the $\Omega$ plane, which, as discussed in Section~\ref{sec-landscape} above, forms the quantum orbits themselves.
These are shown highlighted as curves in \reffigtd{fig-riemann-surface}, and their role here is most clearly apparent in \reffigtdsub{fig-riemann-surface}{fig-riemann-surface-re}: this intersection forms the $(\Re(\omega t), \Re(\Omega))$ energy-time mapping for the quantum orbits, a well-known (and deeply physical) part of the theory~\cite{Ivanov2014, Nayak2019}.
The quantum-orbit Riemann surface, together with its topology, is nothing more than the analytical continuation of this standard energy-time mapping.

\subsection{Derivatives of the action in the two-variable case}
\label{sec-constrained-derivatives}
Before moving on, it is important to define in more detail the relationship between our simple model and the full-blown HHG integral, whose action has two variables instead of one.
This is possible, as mentioned above, by using~\eqref{saddle-point-eqns-ps-tt} to define $t_s'=t_s'(t)$ and with that the single-variable $S_V(t) = S_V(t,t_s'(t))$, but that only works explicitly for the values of the action, and its derivatives need to be considered more carefully.

The first derivative is not affected, since
\begin{align}
\frac{\d S_V}{\d t}(t)
& =
\frac{\d }{\d t} S_V(t,t_s'(t))
\nonumber \\ & = 
\frac{\partial S_V}{\partial t}(t,t_s'(t))
+
\frac{\d t_s'}{\d t}(t)
\frac{\partial S_V}{\partial t'}(t,t_s'(t))
,
\end{align}
and here the partial derivative in the second term, $\frac{\partial S_V}{\partial t'}(t,t_s'(t))$, vanishes by the definition of $t_s'(t)$.
However, if we now turn to the second derivative, the procedure no longer works, and the equivalent calculation,
\begin{align}
\frac{\d^2 S_V}{\d t^2}(t)
& =
\frac{\d }{\d t} \frac{\partial S_V}{\partial t'}(t,t_s'(t))
\nonumber \\ & = 
\frac{\partial^2 S_V}{\partial t^2}(t,t_s'(t))
+
\frac{\d t_s'}{\d t}(t)
\frac{\partial^2 S_V}{\partial t \, \partial t'}(t,t_s'(t))
,
\label{d2Sdt2-partial}
\end{align}
returns a term that includes $\frac{\d t_s'}{\d t}(t)$.
This cannot be evaluated explicitly within this system, as $t_s'(t)$ itself is only defined implicitly.

To resolve this, we use the implicit definition of $t_s'(t)$,
\begin{equation}
\frac{\partial S_V}{\partial t'}(t,t_s'(t)) \equiv 0,
\label{tsp-definition}
\end{equation}
and then differentiate with respect to $t$, to obtain
\begin{align}
\frac{\partial^2 S_V}{\partial t \, \partial t'}(t,t_s'(t))
+
\frac{\d t_s'}{\d t}(t)
\frac{\partial^2 S_V}{\partial t'^2}(t,t_s'(t))
\equiv 0
,
\end{align}
which can then be solved for the implicit derivative as
\begin{align}
\frac{\d t_s'}{\d t}(t)
=
-
\frac{\partial^2 S_V}{\partial t \, \partial t'}(t,t_s'(t))
\bigg/
\frac{\partial^2 S_V}{\partial t'^2}(t,t_s'(t))
.
\label{dts-dt-result}
\end{align}
Finally, this can be substituted into~\eqref{d2Sdt2-partial} to give the form
\begin{align}
\frac{\d^2 S_V}{\d t^2}(t)
& =
\frac{
  \displaystyle
  \frac{\partial^2 S_V}{\partial t^2}
  \frac{\partial^2 S_V}{\partial t'^2}
  -
  \left(
    \frac{\partial^2 S_V}{\partial t \, \partial t'}
    \right)^2
  }{
  \displaystyle
  \frac{\partial^2 S_V}{\partial t'^2}
  }
.
\label{d2Sdt2-final}
\end{align}
Here we have dropped the right-hand side evaluations at $(t,t_s'(t))$ for simplicity, but it is also possible to understand the right-hand side as a function of $(t,t')$ as independent variables, whose zeros are to be found in conjunction with those of~\eqref{saddle-point-eqns-ps-tt}, and this is an easier approach in practice.


Finally, to obtain the higher-order derivatives (in particular, the third derivative, which gives the cubic coefficient $A$) we use symbolic computation~\cite{RBSFA, FigureMaker}, using the substitution rule~\eqref{dts-dt-result} at each step to retain explicit formulations.

\subsection{The threshold harmonics for linear fields}
\label{sec-linear-fields}
For a general driving field, the harmonic cutoffs given by our definition will appear at both ends of the harmonic plateau, oscillating between low frequencies near the ionization threshold and high energies at the classical cutoff.
In general, moreover, all of these harmonic cutoffs will have nonzero imaginary parts, corresponding to missed approaches at a finite distance between the quantum orbits involved.
However, as can be seen in \reffig{fig-dSdt} and \reffig{fig-graph-abs-classified}, this is not the case for the linearly-polarized field we use above, since half of the harmonic-cutoff saddles $t_\hc$ lie directly on the quantum-orbit line, i.e., the quantum orbits have a full crossing there.

This behaviour is generic to all linearly-polarized fields, whether monochromatic or not, and it implies that the low-energy cutoffs always lie at exactly $\Omega_c+i\eta = I_p$, and the threshold harmonics at $\Omega = I_p$ always involve an exact saddle coalescence.
In other words, for threshold harmonics, the saddles of the action landscape are not gaussian: instead, they are exact monkey saddles, as shown in \reffig{fig-saddle-coalescence-Ip} below; this was noticed as early as \citer{Lewenstein1994}, but has not drawn much attention since then.%
\footnote{%
This is largely because the approximations that produced the SFA integral are known to be inaccurate for the lower plateau and threshold harmonics.%
} %

To understand this behaviour, we return to the saddle-point equations~\eqref{saddle-point-eqns-ps}, which for linearly-polarized fields can be rephrased in the simpler, scalar form
\begin{subequations}
\begin{align}
\frac{\partial S_V}{\partial t} 
  = \frac12\left(p_s(t,t')+A(t)\right)^2 + I_p 
& = \Omega
,
\label{saddle-point-eqns-linear-t}
\\
-\frac{\partial S_V}{\partial t'} 
  = \frac12\left(p_s(t,t')+A(t')\right)^2 + I_p
& = 0
.
\label{saddle-point-eqns-linear-tt}
\end{align}
\label{saddle-point-eqns-linear}%
\end{subequations}%
Here the first equation is the crucial one, since at $\Omega=I_p$ it tells us that
\begin{equation}
\frac12 v_r(t,t')^2
  = \frac12\left(p_s(t,t')+A(t)\right)^2 
  = 0
,
\end{equation}
i.e., the recollision velocity $v_r(t,t') = p_s(t,t')+A(t)$ vanishes exactly, with a double zero which still vanishes after taking one derivative.

This becomes particularly important when we consider the second derivative of the action, \eqref{d2Sdt2-final}, whose zeros determine the harmonic-cutoff times. 
The partial derivatives involved in \eqref{d2Sdt2-final} can be evaluated by differentiating \eqref{saddle-point-eqns-linear-t} with respect to $t$ and $t'$, which yields
\begin{subequations}%
\begin{align}
\frac{\partial^2 S_V}{\partial t^2} 
& = \left(p_s+A(t)\right)
    \left[ \frac{\partial p_s}{\partial t} + \dot A(t) \right]
,
\\
\frac{\partial S_V}{\partial t \, \partial t'} 
& = \left(p_s+A(t)\right)
    \frac{\partial p_s}{\partial t'}
.
\end{align}
\label{SV-derivatives-linear}%
\end{subequations}%
Both of these derivatives share a common factor of $v_r(t,t') = p_s(t,t')+A(t)$, and this then means that the second derivative $\frac{\d^2 S_V}{\d t^2}(t)$ will always vanish when evaluated at solutions of the first-order saddle-point equations~\eqref{saddle-point-eqns-linear} at $\Omega=I_p$, completing the proof.

It is important to note that, when a saddle-point coalescence like this occurs, the classification scheme embodied by our test in~\eqref{separatrix-condition} breaks down since, at the double zero, $\frac{\d S_V}{\d t}(t_\hc) $ also vanishes, so the direction vector $\delta t_\sep$ from~\eqref{tsep-definition} also vanishes. 
In practice, this can cause numerical instability, with $\Im\mathopen{}\left[\frac{\d S_V}{\d t}(t_\hc) \right]\mathclose{}$ evaluating to machine precision with a fluctuating sign, and this needs to be handled explicitly -- which could be as simple as assigning an artificial nonzero $\Im\mathopen{}\left[\frac{\d S_V}{\d t}(t_\hc) \right]\mathclose{}$ to be used there, as either direction will work well for the separatrix. 
(For the plots in \reffig{fig-dSdt} and \reffig{fig-graph-abs-classified}, we used a small ellipticity of $\epsilon=0.01\%$ to break the degeneracy and avoid this instability.)

At an exact coalescence, it is impossible to classify the saddles in a unique way: two saddles come in and two saddles come out, but there is no unique way to match either of the outgoing saddles with either of the incoming ones.
This happens for all linearly-polarized fields at the $\Omega=I_p$ threshold, but it can also happen at the harmonic cutoff in isolated cases when the field depends on a variable parameter, as we will exemplify in Section~\ref{sec-bicircular}~below.

\section{The Harmonic-Cutoff Approximation}
\label{sec-yield}
Having examined the structure of the action at the harmonic-cutoff times~$t_\hc$, in this section we will explore the behaviour of the oscillatory integral around it, which we will employ to build the Harmonic-Cutoff Approximation (HCA), an efficient method for estimating the harmonic dipole around the cutoff using only information at~$t_\hc$ itself.

\subsection{Airy-function representation for the model case}
\label{sec-airy-model-qpi}

To do this, we return to the model integral from~\eqref{model-action-5}, where we set as the object of interest the integral the exponential of our model action~\eqref{model-action-3},
\begin{align}
D(\Omega)
=
\int_C e^{-\frac i3 A t^3 +i (\Omega-\Omega_\hc) t} \, \d t
,
\label{D-Omega-Airy-form}
\end{align}
over an integration contour $C$ which should start at the ionization time~$t'$ (or some suitably compatible valley of $\Im(S)$ to the left of~$t_\hc$) and end at $t\to\infty$.
This integral is essentially in Airy form, and it is almost structurally identical to the Airy function's integral representation~\citenisteq{9.5.E4},
\begin{equation}
\Ai(z)
=
\frac{1}{2\pi i}
\int_{\infty e^{-i\pi/3}}^{\infty e^{i\pi/3}}
\exp(\tfrac{1}{3}t^{3}-zt)
\d t
.
\label{Airy-function-integral-representation}
\end{equation}
This is expected, since the harmonic cutoff is a `fold' catastrophe, whose associated diffraction-catastrophe integral is the Airy function~\cite{Berry1980}.

To bring our representation~\eqref{D-Omega-Airy-form} into the canonical form~\eqref{Airy-function-integral-representation}, the core transformation is to eliminate the coefficient in front of the cubic term,
\begin{equation}
-\frac i3 A t^3 \longmapsto \frac13 \tilde t\,^3.
\end{equation}
In a sense, this is relatively simple, as it boils down to a change in integration variable to
$ 
\tilde t = i A^{1/3} \, t,
$ 
but there is an added complication in that the radical $A^{1/3}$ of the cubic coefficient admits three separate branches,
\begin{equation}
\tilde t = i e^{2\pi i k/3} A^{1/3} \, t,
\label{t-tilde-t-transformation}
\end{equation}
for $k\in\{0,1,2\}$, which requires dedicated attention.

\begin{figure}[t]
\setlength{\tabcolsep}{-0.9mm}
\begin{tabular}{ccc}
\subfloat{\label{fig-Airy-contour-zero}
\includegraphics[scale=1]{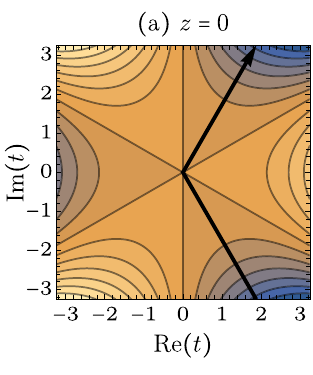} } & 
\subfloat{\label{fig-Airy-contour-real}
\includegraphics[scale=1]{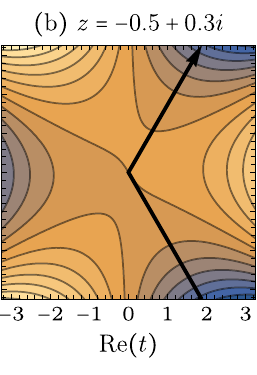} } &
\subfloat{\label{fig-Airy-contour-evanescent}
\includegraphics[scale=1]{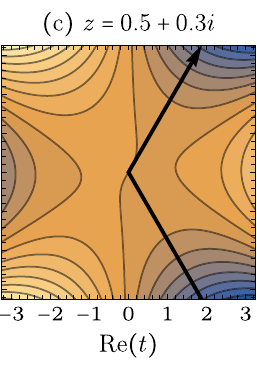} } 
\end{tabular}
\caption{
Contour map of the exponent $\Re(\tfrac13 t^3 -zt)$ of the integral representation of the Airy function,~\eqref{Airy-function-integral-representation}, with the standard contour shown as the black arrows, for 
(a) coalescent saddles,
(b) the oscillatory regime, and
(c) the evanescent part.
The details of the contour can be altered as required (such as e.g. to pass through the two saddle points in~(b) along steepest-descent contours), under the constraints that the contour start at infinity at 
$-\tfrac\pi2 < \arg(t) < -\tfrac\pi6$ 
and end at 
$\tfrac\pi6 < \arg(t) < \tfrac\pi2$,
i.e., that it go down the valleys that surround $e^{-i\pi/3}$ and $e^{i\pi/3}$.
}
\label{fig-Airy-contour-maps}
\end{figure}

The variable change itself is essentially trivial, and \eqref{D-Omega-Airy-form} transforms under~\eqref{t-tilde-t-transformation} to
\begin{subequations}%
\begin{align}
D(\Omega)
& =
\frac{e^{-2\pi i k/3}}{i  A^{1/3}}
\!\!
\int_{\tilde C} \!
\exp(
  \frac 13 \tilde t\,^3
  -\frac{\Omega_\hc-\Omega}{e^{2\pi i k/3} A^{1/3}} \tilde t
  ) 
\,
\d \tilde t 
,
\\ \text{with }
& \tilde C  = i e^{2\pi i k/3} A^{1/3} \,C
.
\end{align}
\end{subequations}%
This is essentially in explicit Airy form, so long as the contour is correct, and thus we can substitute in the Airy function as
\begin{align}
D(\Omega)
& =
\frac{2\pi}{e^{2\pi i k/3}A^{1/3}}
\Ai\mathopen{}\left(
  \frac{\Omega_\hc-\Omega}{e^{2\pi i k/3} A^{1/3}}
  \right)\mathclose{}
,
\label{HCA-result-basic}
\end{align}
where $k$ needs to be chosen such that the altered contour $\tilde C = i e^{2\pi i k/3} A^{1/3} \,C$ is compatible with the standard contour in~\eqref{Airy-function-integral-representation}, which we depict in \reffig{fig-Airy-contour-maps}.

\begin{figure}[t]
\begin{tabular}{c}
\subfloat{\label{fig-saddle-coalescence-Ip}
\includegraphics[scale=1]{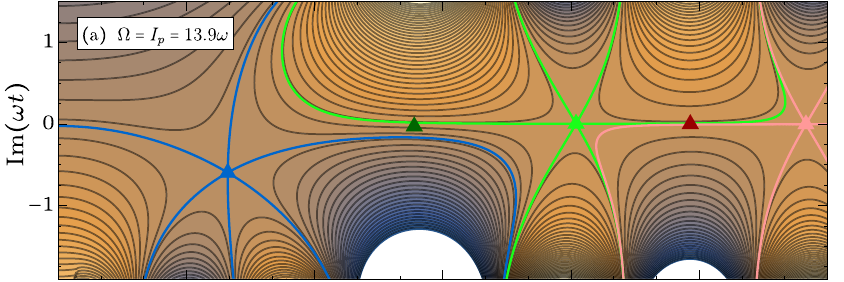} }\\[-2.75mm]
\subfloat{\label{fig-saddle-coalescence-first-hc}
\includegraphics[scale=1]{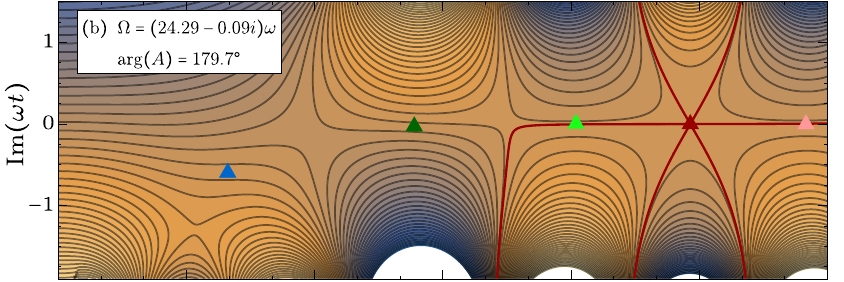} }\\[-2.75mm]
\subfloat{\label{fig-saddle-coalescence-second-hc}
\includegraphics[scale=1]{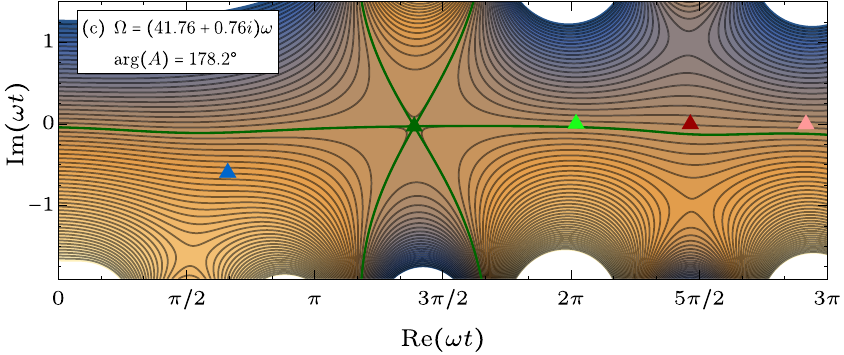} }
\end{tabular}
\caption{
Contour map of $\Im[S(t,t_s'(t))-\Omega t]$, presented as in \reffig{fig-dSdt}, taking a complex $\Omega$ equal to $\Omega_c+i\eta$ at the various harmonic-cutoff times $t_\hc$, causing the nearby saddles 
to fully coalesce, and marking a complete breakdown of the SPA.
To apply the Harmonic-Cutoff Approximation, each monkey saddle (marked by the converging coloured contours) must be rotated by 
$\tilde C = i e^{2\pi i k/3} A^{1/3} \,C$, 
choosing the branch index $k$ so that the integration contour matches the canonical Airy contours of \reffig{fig-Airy-contour-maps}. 
The phase $\arg(A)$ is inset in (b) and (c), and it equals $\SI{-87.97}{\degree}$, $-\SI{0.011}{\degree}$ and $-\SI{0.155}{\degree}$, resp., for the three monkey saddles shown in (a), consistently with their different orientations.
}
\label{fig-saddle-coalescences}
\end{figure}

In principle, a rigorous approach to the contour-choice problem requires a careful examination of the constrains on the contour of the original integral, as shown in \reffig{fig-saddle-coalescences}, to determine the correct integration contour $C$ that is compatible with the original integration limits;
this is then rotated by $i A^{1/3}$ and any additional factors of $e^{2\pi i k/3}$ necessary for the contour to be in canonical form.
(Moreover, care must be taken when taking the radical $A^{1/3}$ over a parameter scan, since, as mentioned in \reffig{fig-saddle-coalescences}, $A$ can lie very close to commonly-used choices for the branch cut of the radical.)

In practice, if there are known constraints on the behaviour of the integral (in particular, exponential decay at $\Omega>\Omega_c$), there will typically only be one choice of $k$ compatible with the constraints, which can then be selected on physical grounds.
As a general rule, exponential decay at $\Omega>\Omega_c$ requires $e^{2\pi ik/3}A^{1/3}$ to have a negative real part.

\begin{figure}[t]
\begin{tabular}{c}
\includegraphics[scale=1]{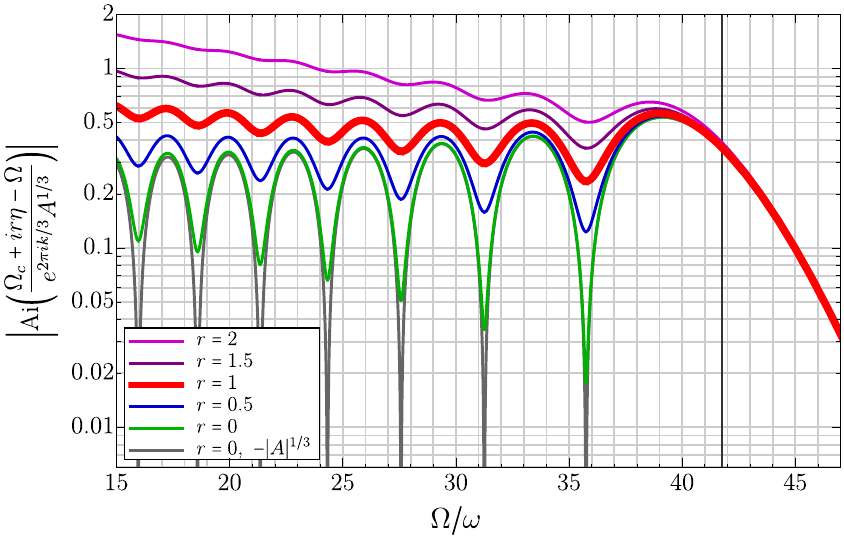}
\end{tabular}
\caption{
Control of the imaginary part of the harmonic cutoff on quantum-path interference.
The red line shows the Airy-function dependence of the Harmonic-Cutoff Approximation, as in~\eqref{HCA-result-basic}, for $\Omega_\hc$ and $A$ corresponding to the first-return cutoff in~\reffig{fig-graphical-abstract}.
The coloured lines have the imaginary part $\eta=\Im(\Omega_\hc)$ artificially reduced or amplified by a variable factor $r$, which respectively amplifies or damps the QPI features;
this corresponds to choosing a lower or higher $\Im(\frac{\partial S_V}{\partial t})$ contour to follow in \reffig{fig-model-contour-map}.
To obtain full interference, $r$ must be set to $0$, but the denominator must also be real and negative, i.e., we replace $e^{2\pi ik/3}A^{1/3}$ with $-|A|^{1/3}$.
The black vertical line is at $\Omega=\Omega_c$.
}
\label{fig-QPI-control}
\end{figure}

Our result for the model integral,~\eqref{HCA-result-basic}, now allows us to have a closer look at the role of the imaginary part of the complex harmonic-cutoff energy $\Omega_\hc$, which we introduced in~\eqref{Im-Omega-hc-setting}.
As we mentioned then, this imaginary part controls the strength of quantum-path interference between the two quantum orbits that meet at the relevant cutoff, and we show this in \reffig{fig-QPI-control}.
The HCA approximates the integral~\eqref{D-Omega-Airy-form} as an explicit Airy function in $\Omega$ (red line), which transitions from oscillatory to exponential-decay behaviour at $\Omega_c=\Re(\Omega_\hc)$.

However, in order for the interference fringes in the oscillatory regime to have a full contrast and pass through the zeros of the Airy function, the argument must be real, and this is generally impossible if $\eta=\Im(\Omega_\hc)$ is nonzero.
To increase the interference features, thus, we can artificially reduce (blue line) or eliminate (green line) this imaginary part; conversely, increasing $\eta$ (purple and magenta lines) further damps the interference, and it introduces exponential behaviour into the plateau.
That said, to get full contrast in the interference it is also necessary for the denominator to be real-valued, which we show as the gray line by artificially adjusting it.

The nonzero imaginary parts of the constants in the argument of the Airy function also makes this approximation distinct from previous approaches that regularize the cutoff in terms of single Airy functions~\cite{Ivanov2014, Frolov2009, Frolov2010}, making it better able to capture the QPI contrast in the neighbourhood of the cutoff.
On the other hand, our approximation sits one level below the Uniform Approximation~\cite{Figueira2002, Milosevic2002, Wong2001} (as well as the equivalent constructions in~Refs.~\citealp{Frolov2012, Sarantseva2013, Okajima2012}), which expands the prefactor to subleading order and is thus able to capture variations in the QPI contrast coming from the prefactor's effect on the saddles.
That said, the HCA is, at least in principle, strictly local to the cutoff, and this makes it especially suited for efficient evaluation and clear analysis of the harmonic strength (as well as phase properties~\cite{Khokhlova2016}) at the cutoff.

\subsection{The full HHG integral}
\label{sec-HCA-full}
We now return to the full two-dimensional integral for HHG,~\eqref{sfa-hhg-dipole-p}, and transform it into the model form~\eqref{D-Omega-Airy-form} so that the HCA can be used to estimate it.
For simplicity, we write the integral in explicit prefactor-action form,
\begin{subequations}%
\begin{align}
\vbD(\Omega)
& =
\int_{-\infty}^\infty \!\!\! \d t
\int_{-\infty}^t \!\!\! \d t' \ 
\vb{f}(t,t') \,
e^{-i S_V\!(t,t') + i\Omega t}
\\
\vb{f}(t,t')
& =
\frac{(2\pi/i)^{3/2}}{(t-t')^{3/2}}
\vbd(\vbp_s(t,t'){+}\vba(t))
\Y(\vbp_s(t,t'){+}\vba(t'))
.
\end{align}
\end{subequations}%
To transform this into the model form~\eqref{D-Omega-Airy-form}, we need to reduce the integral to a single dimension, and then translate the origin to the relevant harmonic-cutoff time $t_\hc$.%
\footnote{%
However, it is important to point out that the process is rather more general than this. 
As was pointed out in the construction of the Uniform Approximation~\cite{Figueira2002} (as well as its earlier analogues in a semiclassical context~\cite{Schomerus1997}), the coordinate separation done here is  in essence an application of the splitting lemma of catastrophe theory~\cite{Poston1978}, which allows the `fold' catastrophe encoded by the Airy function to be isolated into a single coordinate axis.
As such, the simple model of Section~\ref{sec-toy-model} is not a `toy' model in any sense: instead, it is a universal model, which is fully capable of capturing the (local) behaviour of the integral.
} %

The first is achieved, as mentioned earlier, by doing a saddle-point approximation on the $t'$ integral only, and this returns the harmonic dipole in the form
\begin{align}
\vbD(\Omega)
& =
\sum_s
\int_{-\infty}^\infty   \d t \,
\sqrt{\frac{2\pi}{i\frac{\partial^2 S_V}{\partial t'^2}(t,t_s'(t))}}
\vb{f}(t,t_s'(t)) 
\\ \nonumber & \qquad \qquad \qquad \qquad  \times 
e^{-i S_V\!(t,t_s'(t)) + i\Omega t}
,
\end{align}
with $t'_s(t)$ defined implicitly via~\eqref{saddle-point-eqns-ps-tt}. 
After this, we perform a Taylor expansion of the Volkov action $S_V(t,t_s'(t))$ at the harmonic-cutoff time $t_\hc$ up to third order,
\begin{align}
\vbD(\Omega)
& =
\sum_{s,\hc}
\int_{C} \!\! \d t 
\sqrt{\frac{2\pi/i}{\frac{\partial^2 S_V}{\partial t'^2}(t,t_s'(t))}}
\vb{f}(t,t_s'(t)) 
e^{-i S_V\!(t_\hc,t_s'(_\hc))}
\\ \nonumber & \qquad  \times 
\exp(
-\frac i3 A (t-t_\hc)^3
-i \Omega_\hc (t-t_\hc)
+ i\Omega t
)
,
\end{align}
where $A$ and $\Omega_\hc$ are the third and first derivatives of $S_V(t,t_s'(t))$, as set in~\eqref{action-derivatives-thc} and evaluated as constructed in Section~\ref{sec-constrained-derivatives};
here we have also allowed the integration contour $C$ to vary so that it can pass through $t_\hc$ as appropriate.
Finally, we move the integration origin to $t_\hc$, 
\begin{align}
\vbD(\Omega)
& =
\sum_{s,\hc}
\int_{\bar C} \!\! \d \bar t 
\sqrt{\frac{2\pi/i}{\frac{\partial^2 S_V}{\partial t'^2}(t,t_s'(t))}}
\vb{f}(t,t_s'(t)) 
e^{-i S_V\!(t_\hc,t_s'(_\hc))}
\\ \nonumber & \qquad  \times 
\exp(
-\frac i3 A \bar t\,^3
-i \Omega_\hc \bar t
+ i\Omega (\bar t + t_\hc)
)
,
\end{align}
keeping the explicit $t=\bar t+t_\hc$ for notational simplicity.

To apply the Harmonic-Cutoff Approximation, we now assume that the prefactor varies slowly enough at $t_\hc$ that it can be pulled out of the integral (i.e., that it changes slower than the decay into the valleys shown in \reffig{fig-saddle-coalescences}),%
\footnote{%
On a more rigorous footing, this amounts to a zeroth-order Taylor expansion of the prefactor at $t_\hc$, as is done in the SPA~\cite{GerlachSPAonline}. 
Derivatives of the prefactor can be added as required, and the corresponding terms will change the Airy function to its derivatives 
(as can be seen by differentiating~\eqref{Airy-function-integral-representation} with respect to $z$, which brings down a factor of $t$ into the prefactor).
For the cases we plot here, these terms are about two orders of magnitude weaker than the leading-order contribution.
} %
which gives
\begin{align}
\vbD(\Omega)
& =
\sum_{\hc}
\sqrt{\frac{
    2\pi
    }{
    i\frac{
       \partial^2 S_V
       }{
       \partial t'^2
       }(t_\hc^{\phantom a},t'_\hc)
    }}
\vb{f}(t_\hc^{\phantom a},t'_\hc)
e^{-i S_V\!(t_\hc^{\phantom a},t_\hc')}
\\ \nonumber & \qquad  \times 
e^{i\Omega t_\hc}
\int_{\bar C} 
\exp(
-\frac i3 A \bar t\,^3
+ i(\Omega-\Omega_\hc)\bar t 
)
\d \bar t 
.
\end{align}
The integral is now in the form of~\eqref{D-Omega-Airy-form}, with an added functional dependence on $\Omega$ coming from the term $e^{i\Omega t_\hc}$.
This term results from the translation of the time origin, and it can have a sizeable effect on the harmonic yield if $\Im(t_\hc)$ is nonzero.
That said, we can now use the result~\eqref{HCA-result-basic} for the model form directly.

We obtain thus our final result for the HCA,
\begin{align}
\vbD(\Omega)
& =
\sum_{\hc}
\sqrt{\frac{
    2\pi
    }{
    i\frac{
       \partial^2 S_V
       }{
       \partial t'^2
       }(t_\hc^{\phantom a},t'_\hc)
    }}
\frac{2\pi}{e^{2\pi i k/3}A_\hc^{1/3}}
\\ \nonumber & \quad  \times 
\vb{f}(t_\hc^{\phantom a},t'_\hc)
e^{-i S_V\!(t_\hc^{\phantom a},t_\hc')+i\Omega t_\hc}
\Ai\mathopen{}\left(
  \frac{\Omega_\hc-\Omega}{e^{2\pi i k/3} A_\hc^{1/3}}
  \right)\mathclose{}
,
\end{align}
where the harmonic-cutoff time pairs $(t_\hc^{\phantom a},t'_\hc)$ are found by solving the simultaneous equations
\begin{subequations}%
\begin{align}
\frac{\partial^2 S_V}{\partial t^2}
\frac{\partial^2 S_V}{\partial t'^2}
-
\left(
  \frac{\partial^2 S_V}{\partial t \, \partial t'}
  \right)^2
& =
0
\label{thc-equations-recap-d2S}
\\
\frac{\partial S_V}{\partial t'}
& = 0
\label{thc-equations-recap-tt}
\end{align}
\label{thc-equations-recap}%
\end{subequations}%
and where the coefficients $\Omega_\hc$ and $A_\hc$ for each pair of harmonic-cutoff times $(t_\hc^{\phantom a},t'_\hc)$ are given, as in \eqref{action-derivatives-thc}, by
\begin{align}
\Omega_\hc = \frac{\partial S_V}{\partial t} (t_\hc^{\phantom a},t'_\hc)
\ \text{and} \ 
A_\hc = \frac12 \frac{\d^3 S_V}{\d t^3} (t_\hc^{\phantom a},t'_\hc)
,
\end{align}
with the third derivative $\frac{\d^3}{\d t^3}$ understood in the constrained sense of Section~\ref{sec-constrained-derivatives}.%
\footnote{%
An added wrinkle can appear if the prefactor, and particularly the dipole moments it includes, have singularities at the solutions of~\eqref{thc-equations-recap}, which is often the case in the regular SPA.
If this occurs, a regularization scheme like the one described in \citer{Popruzhenko2014} will be required, but this is likely to extend relatively cleanly.
} %
Our implementation of this approximation, in the Wolfram Language, is available in the RBSFA software package~\cite{RBSFA}.

\begin{figure}[t]
\begin{tabular}{c}
\subfloat{\label{fig-HCA-results-action}
\includegraphics[scale=1]{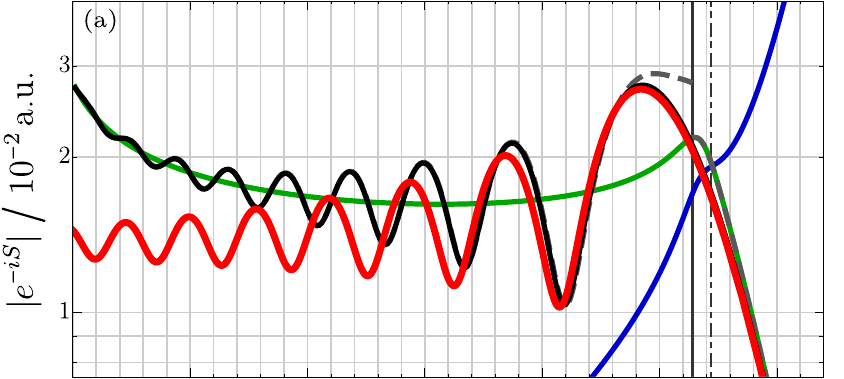} }\\[-2.5mm]
\subfloat{\label{fig-HCA-results-action-tau}
\includegraphics[scale=1]{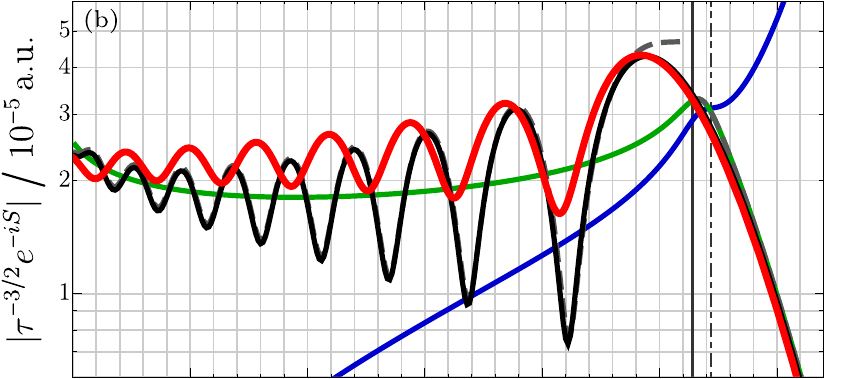} }\\[-2.5mm]
\subfloat{\label{fig-HCA-results-full-dipole}
\includegraphics[scale=1]{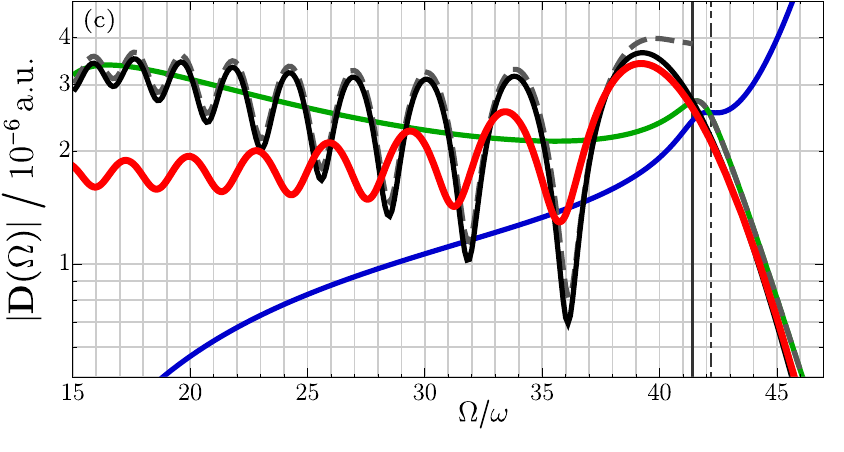} }
\end{tabular}
\vspace{-7mm}
\caption{
Approximations to 
(a) the pure action term $e^{-iS}$, 
(b) the action term with wavepacket diffusion, $\tau^{-3/2}e^{-iS}$, and
(c) the full harmonic dipole, 
via
the Saddle-Point Approximation for the long and short trajectories (green and blue lines) and
their coherent combination (dashed gray lines),
the Uniform Approximation (black lines), and
the Harmonic-Cutoff Approximation (red lines).
The HCA is quantitatively accurate at the harmonic cutoff, and reasonably qualitatively accurate in the upper plateau, but it only requires solving a single saddle-point equation, instead one per harmonic frequency over a tight grid in $\Omega$.
}
\label{fig-HCA-results}
\end{figure}

We show the results of this approximation, compared to the SPA and the UA (as described in Refs.~\citealp{Figueira2002, Milosevic2002}), in \reffig{fig-HCA-results}, plotted as in \reffig{fig-traj-approach-spectra}.
When the prefactor is fully ignored, as in \reffig{fig-HCA-results-action}, the HCA is extremely accurate at the harmonic cutoff, and it retains a good qualitative accuracy as $\Omega$ descends from~$\Omega_c$ and into the plateau: it shifts vertically from the UA in the lower plateau and the predicted period for the QPI beatings is too long, but the QPI contrast is mostly well reproduced.

The qualitative accuracy of the HCA, particularly regarding the QPI contrast, gets degraded when the wavepacket-diffusion dilution factor of $\tau^{-3/2}$ is included, as shown in \reffig{fig-HCA-results-action-tau}.
This happens because the $\tau^{-3/2}$ factor affects the long trajectories (green line) much more than it does the short ones (blue line), bringing them closer together and allowing for better interference between them;
the HCA, with its information coming from a single point, is blind to this effect, which only acts as a global vertical shift on the curve. 
The same is true for the full harmonic dipole $\vbD(\Omega)$,%
\footnote{%
Here we use the ground-state wavefunction and dipole transition matrix element for the $s$ state of a short-range-potential, as described in \citer{Pisanty2017}.
} %
which we plot in \reffig{fig-HCA-results-full-dipole} -- the HCA has good quantitative accuracy at the cutoff, but it becomes more of a qualitative estimation for the middle and lower plateau.

However, this weakness of the HCA is also its biggest strength: 
precisely because it is able to describe the harmonic yield using only information coming from a single solution of a set of saddle-point equations \eqref{thc-equations-recap}, it can be calculated in a small fraction of the time taken to produce SPA or UA calculations of the harmonic spectrum, since those require the solution of one system of saddle-point equations -- the system in~\eqref{saddle-point-eqns-ps} -- for each harmonic frequency~$\Omega$ of interest, and these will typically form a grid with hundreds of instances.

In this sense, the HCA sits at the far end of the tradeoff spectrum between numerical accuracy and computational complexity, which is typically understood as running from full simulations of the time-dependent Schr\"odinger equation (TDSE)~\cite{Scrinzi2014}, passing through explicit numerical integration of the SFA dipole~\eqref{sfa-hhg-dipole-p}, to the SPA and UA quantum-orbit approaches.
Normally, the quantum-orbit methods are considered to be fast enough that no further optimization is necessary, but this is not always the case when large scans are required spanning a high-dimensional parameter space---particularly if the waveforms involved cause nontrivial behaviour in the quantum orbits---, such as when optimizing the length or strength of the harmonic plateau and cutoff~\cite{Chipperfield2009, Chipperfield2010, Haessler2014}.
Moreover, when the primary focus is on the harmonic cutoff, the HCA will typically be sufficiently accurate.

\section{Parameter scaling and the cutoff law}
\label{sec-scaling}
As we have seen, the formalism we have presented allows us to extract both a unique and natural definition for the (complex) times that correspond to the harmonic cutoff, and also a natural definition of the harmonic frequency $\Omega_c$ at which the cutoff occurs.
This identification is a valuable connection, since the high-harmonic cutoff is often understood as a vague term, referring to a spectral region where the behaviour changes, instead of a concrete point---largely because of the difficulty in pinning down a specific frequency at which this change occurs.

This is particularly important, since the scaling of the cutoff frequency with the field parameters is one of the key hallmarks of HHG, in terms of the so-called `cutoff~law',
\begin{equation}
\Omega_\mathrm{max}
\approx
3.17 \, U_p + I_p 
,
\phantom{\, F(I_p/U_p)}
\label{scaling-law-classical}
\end{equation}
which relates the cutoff frequency $\Omega_\mathrm{max}$ to the target's ionization potential $I_p$ and the ponderomotive energy $U_p=F^2/4\omega^2$ of the field.
This relationship was first uncovered in numerical simulations~\cite{Krause1992} and subsequently explained using the classical three-step model~\cite{Corkum1993, Kulander1993}: the numerical factor of $3.17$ arises as the maximal kinetic energy that can be achieved at the return to the origin by an electron released at zero velocity, with the $I_p$ term representing quantum energy conservation at the recombination, with an essentially heuristic justification.

The fully-quantum quasiclassical theory~\cite{Lewenstein1994} re-derives this cutoff law, giving also a quantum correction to the $I_p$ term of the form
\begin{equation}
\Omega_\mathrm{max}
\approx
3.17 \, U_p + 1.32 \, I_p
.
\nhphantom{1.32\,}
\phantom{\, F(I_p/U_p)}
\label{scaling-law-quantum-corrections}
\end{equation}
To obtain this formulation, the cutoff frequency is defined as the maximum of the recollision kinetic energy, $\Re(\vbv(t_r)^2)$, taken under the restriction that its imaginary part $\Im(\vbv(t_r)^2)$ vanish.
The dynamics are then analyzed using a systematic expansion on $I_p/U_p$, using the purely-classical case at $I_p=0$ for the leading $3.17U_p$ term, after which the cutoff law can be derived in the exact form
\begin{equation}
\Omega_\mathrm{max}
=
3.17 \, U_p + F(I_p/U_p)\, I_p
,
\label{scaling-law-quantum-corrections-F}
\end{equation}
in terms of a universal function $F(I_p/U_p)$ which can be calculated numerically as $F(0) = 1.32$.

This prescription can be computed exactly for linearly-polarized monochromatic fields, and it can be extended to elliptical polarizations~\cite{Milosevic2000, Flegel2004}, but it is laborious to apply for polychromatic combinations~\cite{Figueira1999} (where the complex waveform makes analytical calculations challenging) and for tailored field polarizations~\cite{Milosevic1996}, especially for field shapes whose recolliding orbits do not ionize at zero velocity, as required by the classical theory.
In any case, this definition of the harmonic cutoff has not seen wide adoption in the literature.

Instead, a wide variety of other methods have been used in its place, including 
relaxing the $\Im(\vbv(t_r)^2)=0$ condition to $\Im(t_r)=0$~\cite{Milosevic2000bicircular},
focusing only on the classical recollision velocity~\cite{Chipperfield2009, Chipperfield2010},
examining the change in the intensity scaling of the harmonic yield at a fixed harmonic order~\cite{Lewenstein1995},
the use of graphical methods based on tangents to the optical waveform~\cite{Figueira1999},
and extracting the cutoff from the Bessel-function expansion of the oscillatory integral~\cite{Averbukh2001, Avetissian2014},
as well as direct numerical methods used to extract the cutoff from the results of TDSE simulations~\cite{Neufeld2019timescales}.
None of these definitions, however, is particularly satisfactory, and -- like the original definition -- none of the analytical approaches have been used very widely in the literature.

\begin{figure}[b]
\begin{tabular}{c}
\subfloat{\label{fig-scalings-re-Omega}
\includegraphics[scale=1]{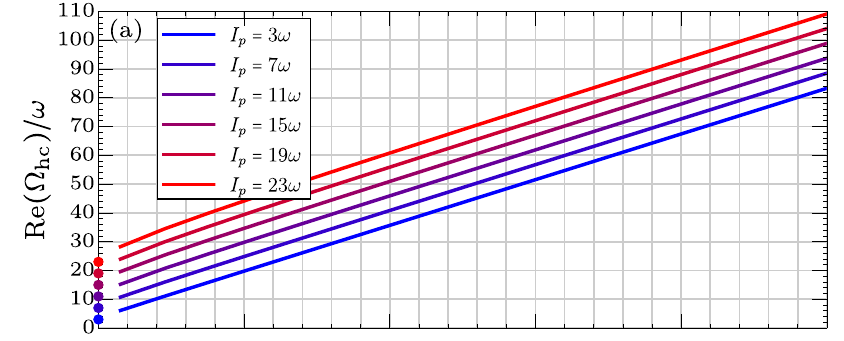} }\\[-3.5mm]
\subfloat{\label{fig-scalings-im-Omega}
\includegraphics[scale=1]{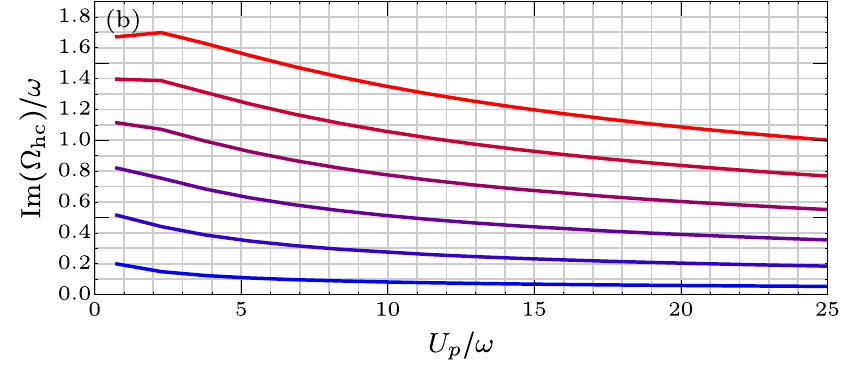} }\\[-0.8mm]
\subfloat{\label{fig-scalings-re-F}
\includegraphics[scale=1]{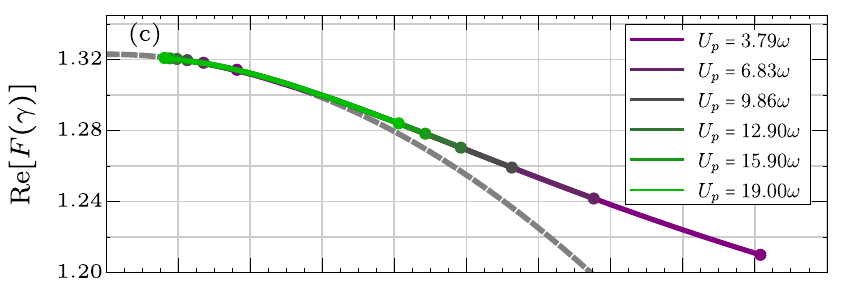} }\\[-3.5mm]
\subfloat{\label{fig-scalings-im-F}
\includegraphics[scale=1]{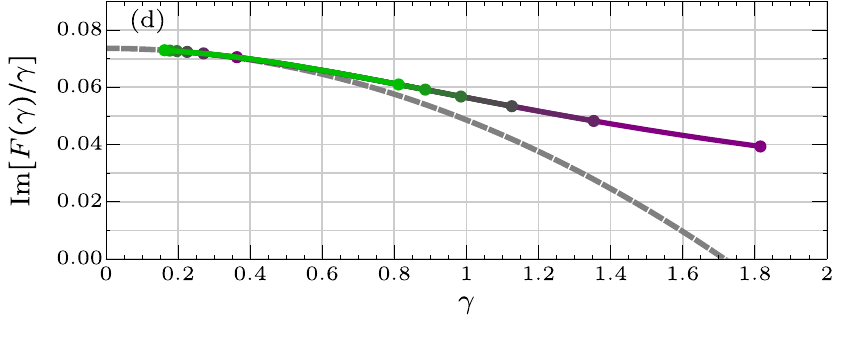} }
\end{tabular}
\caption{
Scaling of the harmonic-cutoff frequency $\Omega_\hc$ with respect to the field parameters.
(a)~The affine dependence of the cutoff law $\Re(\Omega_\hc) \approx 3.17 U_p + I_p$ is reproduced well; the dots at the axis mark the position of $I_p$, and the mismatch to the lines comes from the quantum corrections.
(b)~The imaginary part $\Im(\Omega_\hc)$ decays with $U_p$ as $I_p^{3/2}/U_p^{1/2}$.
These scalings indicate a relationship of the form $\Omega_\hc = 3.17U_p + F(\gamma) I_p$, 
with the real and imaginary parts of $F(\gamma)$, shown in (c) and (d), suggesting that it is a power series in $i\gamma$.
}
\label{fig-scalings}
\end{figure}

As we argued earlier, the real part $\Omega_c$ of our complex-valued $\Omega_\hc$ forms a natural candidate as a precise definition of the harmonic-cutoff frequency $\Omega_\mathrm{max}$, with the added advantage that it can be trivially adapted to complicated waveforms and to tailored polarizations without significantly complicating the calculation, a useful feature as the complexity of the field shapes under consideration continues to increase~\cite{Neufeld2019, Javsarevic2020}. 
To strengthen this identification, though, it is important to verify that it reproduces the cutoff law, including the quantum corrections coming from the SFA as captured by the original definition.
We show this in \reffig{fig-scalings-re-Omega}: the variation of $\Re(\Omega_\hc)$ with $U_p$ is linear with an $I_p$-dependent offset, which closely follows the quasiclassical quantum-corrected cutoff law~\eqref{scaling-law-quantum-corrections}.

However, our formalism allows us to go beyond this level, since $\Omega_\hc$ also has an imaginary part, whose scaling with $U_p$ and $I_p$ is shown in \reffig{fig-scalings-im-Omega}.
As rough behaviour, $\Im(\Omega_\hc)$ decreases with $U_p$ and increases with $I_p$, and simple testing shows that the leading asymptotic component is the behaviour
\begin{equation}
\Im(\Omega_\hc)
\sim
{I_p^{3/2}} \big/ {U_p^{1/2}}
.
\end{equation}
At first glance, this shows very different behaviour to the real part, which follows~\eqref{scaling-law-quantum-corrections-F}.
However, once a factor of $I_p$ has been set apart, as we did for the real part, the scaling for $\Im(\Omega_\hc)$ is simply the Keldysh parameter, $\gamma = \sqrt{I_p/2U_p}$.
This is therefore indicative that the parameter in the quantum scaling law~\eqref{scaling-law-quantum-corrections-F} should be amended from $I_p/U_p$ to its square root, so it should read
\begin{equation}
\Omega_\hc
=
3.17 \, U_p + F(\gamma)\, I_p
.
\label{scaling-law-quantum-F-gamma}
\end{equation}

In this form,~\eqref{scaling-law-quantum-F-gamma} essentially acts as a definition for $F(\gamma)=(\Omega_\hc - 3.17U_p)/I_p$, but this definition can only work if that value is independent of what combination of $I_p$ and $U_p$ gives rise to $\gamma$.
This is indeed the case, as we show in Figs.~\ref{fig-scalings-re-F}
and~\ref{fig-scalings-im-F} for the real and imaginary parts, respectively: plotting $F(\gamma)$ by changing $I_p$ reveals the same universal curve for a range of different values of $U_p$.
Moreover, we can extract the low-$\gamma$ behaviour here to obtain
\begin{equation}
\begin{aligned}
F(\gamma) 
& \approx 
1.323 + i \, 0.07361 \, \gamma 
\\ & \quad \ 
- 0.068 \, \gamma^2 - i \, 0.025 \, \gamma^3 + \cdots
,
\end{aligned}
\end{equation}
with the first two terms shown as the gray dashed line in Figs.~\ref{fig-scalings-re-F} and~\ref{fig-scalings-im-F}.
This understanding coincides with the existing theory (so, in particular, \reffig{fig-scalings-re-F} coincides with Fig.~5 of \citer{Lewenstein1994} when plotted as a function of $\gamma^2$), but it also helps extend it and clarify its structure, though a formal analysis of the existence and convergence of the power series in $i\gamma$ for $F(\gamma)$, as defined here for $\Omega_\hc$, is still lacking.
We summarize our parameter-scaling results, and their relationship to the existing theories, in Table~\ref{table-scalings-summary}.

\begin{table}[h]
\begin{tabular}{rcl}
Theory & \hspace{9pt} & Cutoff law
\\ \addlinespace[0.5mm] \toprule[0.1em] \addlinespace[1mm]

Corkum~\cite{Corkum1993} & &
\multirow{2}{*}{  \hspace{-1mm}$\Omega_\mathrm{max} \approx 3.17 U_p + I_p$} \\
Kulander~\cite{Kulander1993} & &

\\ \addlinespace[1mm] \hline \addlinespace[1mm]
Lewenstein~\cite{Lewenstein1994} & &
$\begin{aligned}
\Omega_\mathrm{max} & = 3.17 U_p + F(I_p/U_p) I_p \\
F(0) & = 1.32
\end{aligned}$

\\ \addlinespace[1mm] \hline \addlinespace[1mm]
This work & &
$\begin{aligned}
\Omega_\hc & = 3.17 U_p + F(\gamma) I_p \\
F(\gamma) & = 1.323 + i \, 0.07361 \, \gamma 
\\ & \quad \ 
- 0.068 \, \gamma^2 - i \, 0.025 \, \gamma^3 + \cdots
\end{aligned}$

\\ \addlinespace[1mm] \hline
\end{tabular}

\caption{
Scaling of the harmonic cutoff with $I_p$ and $U_p$, for the different understandings of the cutoff.
}
\label{table-scalings-summary}
\end{table}

\section{Nontrivial quantum-orbit topology for bicircular fields}
\label{sec-bicircular}
To conclude our (current) explorations of the role of the complex harmonic-cutoff times in HHG, we return to the classification scheme for separating saddle-point solutions into quantum orbits, which we first showcased in \reffig{fig-graphical-abstract} and constructed in detail in Section~\ref{sec-saddle-orientation}.
One crucial aspect of this classification scheme is the orientation of the separatrix, which indicates the direction that the saddle points take in their avoided crossing at the cutoff: i.e., whether the short-trajectory saddle, coming in from the left, will go up into positive (or down into negative) imaginary time when it meets the long-trajectory saddle.
Our construction shows that this direction is given, via~\eqref{tsep-definition},~by 
\begin{align}
\delta t_\sep = \sqrt{-i\eta/A_\hc}
,
\end{align}
and thus that it has a sensitive dependence on the sign of 
\begin{equation}
\eta = \frac{\partial S_V}{\partial t}(t_\hc^{\phantom a},t_\hc')
,
\end{equation}
the imaginary part of the time derivative of the Volkov action at the harmonic-cutoff time.

This imaginary part is essentially controlled by the internal details of the action and, as such, its sign can be either positive or negative depending on the precise particulars of the optical waveform of the driving laser.
If the shape of the driving laser pulse remains essentially constant (say, when doing an intensity scan as in \reffig{fig-scalings}) then the sign of the imaginary part is unlikely to change (as was seen to be the case in \reffig{fig-scalings-im-Omega}).

However, if one has a more complicated driving field with a nontrivial shape parameter that affects the waveform of the pulse, and thus the details of the Volkov action, then the sign of $\eta$ will change, and the direction of the separatrix, $\delta t_\sep$, will switch to one $\SI{45}{\degree}$ away.
This will, in turn, switch the direction of the avoided crossing, and with that the identity (say, as `short' or `long' trajectories) of the post-cutoff branches of the orbit.
This behaviour is generic and commonplace: as a simple example, it can be observed in monochromatic fields once an envelope is introduced, where changes in the carrier-envelope phase can produce this type of topological change in the quantum-orbit layout.

In this section we showcase a concrete example of this behaviour, and we explore the nontrivial topologies that it induces in the set of quantum-orbit saddle points. 
We focus, in particular, on tailored polarization states known as `bicircular' fields~\cite{Fleischer2014, Kfir2015, Eichmann1995, Milosevic1996, Milosevic2000bicircular, Baykusheva2016, Dorney2017, JimenezGalan2018, Pisanty2014, Pisanty2017, Milovsevic2019}, formed by the combination of two counter-rotating circularly-polarized strong laser pulses at frequencies $\omega$ and $2\omega$,
\begin{equation}
\vbf(t)
=
F_1 \begin{pmatrix} \cos(\omega t) \\ \sin(\omega t) \end{pmatrix}
+
F_2 \begin{pmatrix} \phantom{-}\cos(2\omega t) \\ -\sin(2\omega t) \end{pmatrix}
.
\end{equation}
These fields have been the focus of widespread interest in recent years because they are subject to a spin selection rule~\cite{Fleischer2014, Pisanty2014} which ensures that the harmonics of the driver are circularly polarized~\cite{Kfir2015}.
For our purposes, however, we select them as an example of a reasonably well-understood waveform with a nontrivial shape parameter, namely, the relative intensity between the two pulses; this is normally kept at unity, but the effects of its variation have seen some exploration~\cite{Milosevic2000bicircular, Baykusheva2016, Dorney2017, JimenezGalan2018}.
If we keep the total intensity constant, at $I_\mathrm{tot} = \SI{2e14}{W/cm^2}$ for concreteness, then the shift in intensity is best described by the mixing angle $\theta$, which we use to define the individual field amplitudes as
\begin{align}
F_1 = F \cos(\theta),
\qquad
F_2 = F \sin(\theta)
,
\end{align}
where $F=\sqrt{2}\times\SI{0.053}{\au}$ is the peak field strength.

\begin{figure}[t]
{
\setlength{\tabcolsep}{-0.5mm}
\newlength{\figureKspacing}
\setlength{\figureKspacing}{-3mm}
\begin{tabular}{c}
\subfloat{\label{fig-bicircular-transitions-detail-a} %
\includegraphics[scale=1]{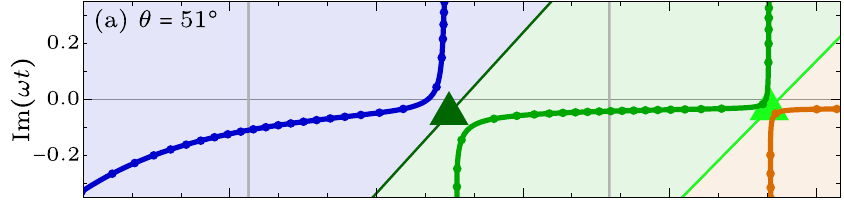} }\\[\figureKspacing]
\subfloat{\label{fig-bicircular-transitions-detail-b} %
\includegraphics[scale=1]{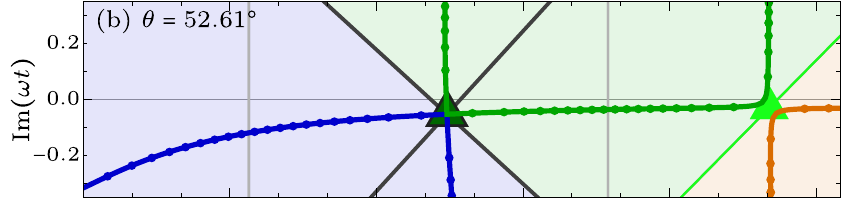} }\\[\figureKspacing]
\subfloat{\label{fig-bicircular-transitions-detail-c} %
\includegraphics[scale=1]{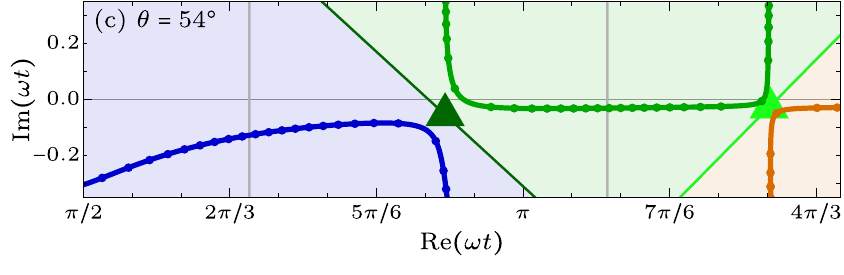} }
\end{tabular}
}
\caption{
Topological transition and reconnection in the quantum orbits of a bicircular field as the mixing angle~$\theta$ changes.
Here we show a detail of the quantum-orbit layout for the short and long trajectories, plotted as in \reffig{fig-graph-abs-classified}, with the triangles marking the harmonic-cutoff times $t_\hc$ and the coloured separatrices marking the regions occupied by the different orbits, as described in Section~\ref{sec-saddle-orientation}.
In~(b) we show the topological transition, where $\eta=\Im(\tfrac{\d S_V}{\d t})$ vanishes and the separatrix becomes undefined, which we show by 
 highlighting the $t_\hc$ with a black border and
 plotting both options (using an artificial $\eta=\pm 1$) as the gray cross.
In this case, both choices of separatrix are acceptable for saddle classification, since the saddles fully coalesce.
}
\label{fig-bicircular-transitions-detail}
\end{figure}

We show in \reffig{fig-bicircular-transitions-detail} a representative topological reconnection transition, between the short and long orbits (blue and green curves, respectively) of the bicircular field.
Before the transition, in~\reffig{fig-bicircular-transitions-detail-a}, the short orbit connects up to the positive-imaginary saddle, while after the transition, in~\reffig{fig-bicircular-transitions-detail-c}, it connects to negative imaginary time.
Between these two there must always lie a transition, which we show in~\reffig{fig-bicircular-transitions-detail-b}:
here the saddles fully coalesce, producing a monkey-saddle landscape like the ones shown in \reffig{fig-saddle-coalescences}, with a purely-real HCA Airy function corresponding to enhanced quantum path interference between the two orbits.

\begin{figure*}
{
\setlength{\tabcolsep}{-0.5mm}
\newlength{\figureJspacing}
\setlength{\figureJspacing}{-3mm}
\begin{tabular}{rl}
\begin{tabular}{r}
\subfloat{\label{fig-bicircular-transitions-a} %
\includegraphics[scale=1]{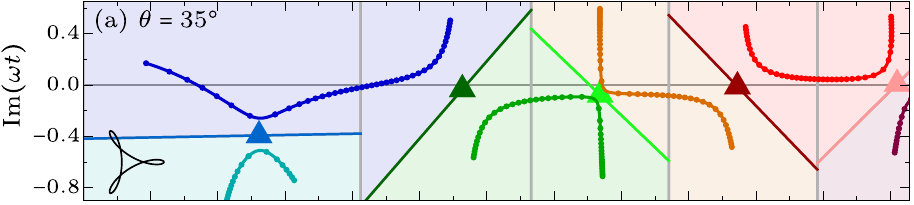} } \\[\figureJspacing]
\subfloat{\label{fig-bicircular-transitions-b} %
\includegraphics[scale=1]{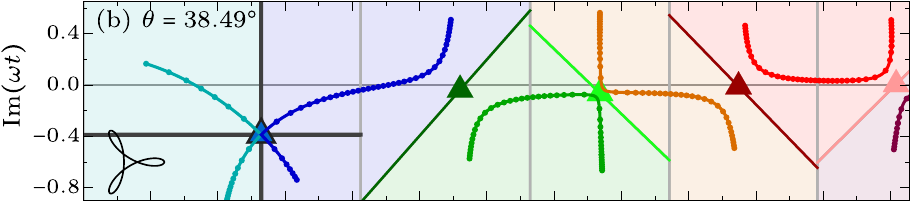} } \\[\figureJspacing]
\subfloat{\label{fig-bicircular-transitions-c} %
\includegraphics[scale=1]{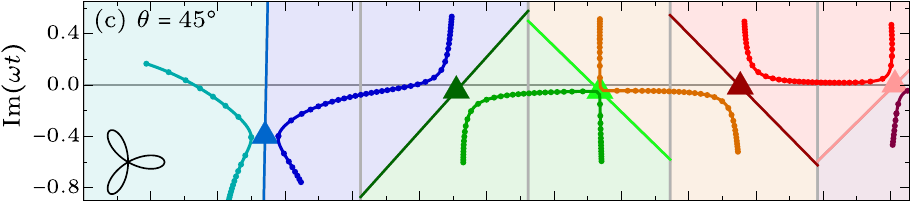} } \\[\figureJspacing]
\subfloat{\label{fig-bicircular-transitions-d} %
\includegraphics[scale=1]{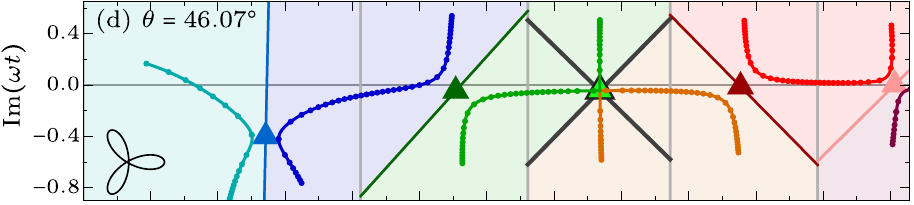} } \\[\figureJspacing]
\subfloat{\label{fig-bicircular-transitions-e} %
\includegraphics[scale=1]{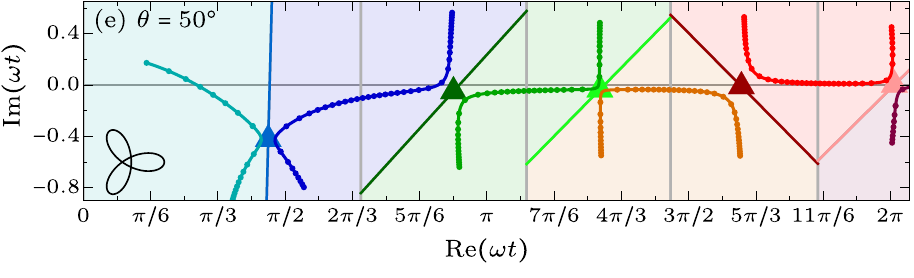} } 
\end{tabular}
&
\begin{tabular}{r}
\subfloat{\label{fig-bicircular-transitions-f} %
\includegraphics[scale=1]{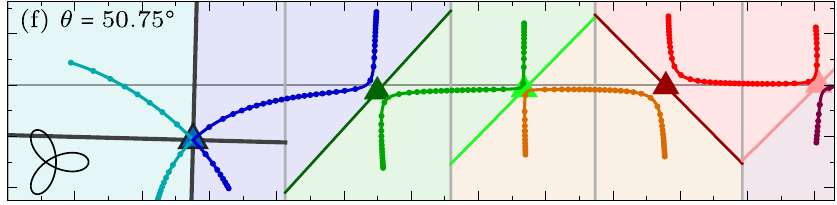} }\\[\figureJspacing]
\subfloat{\label{fig-bicircular-transitions-g} %
\includegraphics[scale=1]{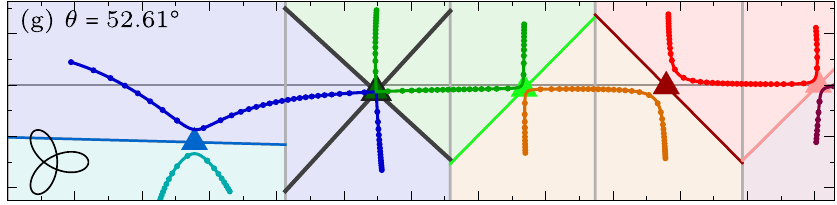} }\\[\figureJspacing]
\subfloat{\label{fig-bicircular-transitions-h} %
\includegraphics[scale=1]{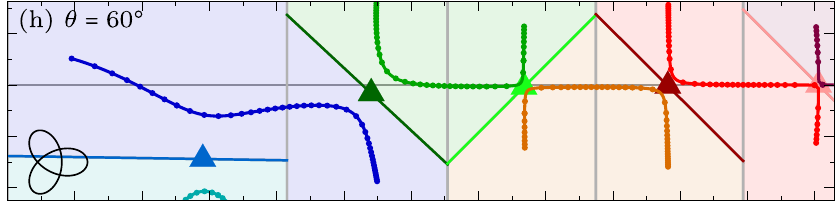} }\\[\figureJspacing]
\subfloat{\label{fig-bicircular-transitions-i} %
\includegraphics[scale=1]{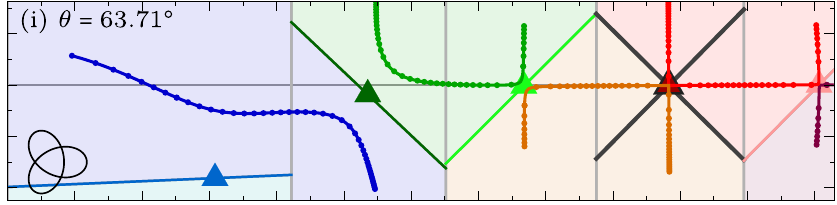} }\\[\figureJspacing]
\subfloat{\label{fig-bicircular-transitions-j} %
\includegraphics[scale=1]{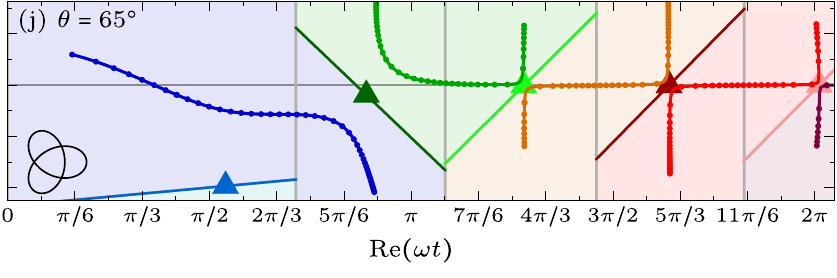} }
\end{tabular}
\end{tabular}
}
\caption{%
Topological transitions and reconnections in the quantum orbits of a bicircular field as a function of the mixing angle~$\theta$, plotted as in \reffig{fig-bicircular-transitions-detail}.
As $\theta$ varies over a wider range, the various pairs of neighbouring quantum orbits go through reconnection transitions with each other (with two separate transitions, in~(b) and~(f), for the first pair of orbits), so that all of the quantum orbits are connected into a single topology, which we show below in \reffigtd{fig-topology-3D}.
At each panel the inset at bottom right shows the shape of the electric field at the relevant mixing angle.
}
\label{fig-bicircular-transitions}
\end{figure*}

More importantly, however, this transition separates the instances with $\eta>0$ from those with $\eta<0$, so at the transition itself $\eta$ must vanish -- and, as with the exact coalescences we studied in Section~\ref{sec-linear-fields}, no (nonzero) unique separatrix direction $\delta t_\sep$ can be found.
That said, this failure is fairly benign since, at the transition, either of the separatrices obtained by the artificial choices of $\eta = \pm 1$ will work correctly.
At the transition the missed approach becomes a full coalescence, and any identification of pre- and post-cutoff orbits is artificial, so both signs of $\eta$ will work equally well.
In \reffig{fig-bicircular-transitions-detail-b}, we mark the transition by showing both possible separatrices in gray, taking an arbitrary choice between the two as to which one to use for the actual classification.
We also highlight the triangle at the harmonic-cutoff time~$t_\hc$ with a black border for further clarity.

On a more practical footing, the fact that $\eta=\Im(\tfrac{\d S_V}{\d t})$ vanishes at the transition is especially useful, since it can be used to look for the parameters where the transitions happen, by looking for zeros of $\eta$ as a function of~$\theta$.

That said, the most important aspect of these topological transitions becomes apparent when we survey the quantum-orbit landscape on a wider perspective, in terms of the multiple quantum orbits present in the dynamics,  as well as for a wider interval in the mixing angle, which we show in \reffig{fig-bicircular-transitions}.
Here we see that all of the pairs of quantum orbits undergo one or more reconnection transitions with each other, so that no pair of quantum orbits, however apparently distant, can actually be separated: 
in the same way that the evanescent orbits of \reffig{fig-bicircular-transitions-detail-b} have `confused' identities between the short and long orbits,
the short orbit `mingles' at low $\Omega$ with the short-time orbit shown in cyan, at two separate transitions (shown in~\reffig{fig-bicircular-transitions-b} and~\reffig{fig-bicircular-transitions-f}),
the long orbit connects with the second return (with the transition shown in \reffig{fig-bicircular-transitions-d}), 
and so on.

\begin{tdfigure*}
\begin{tabular}{c}
\href{https://imaginary-harmonic-cutoff.github.io/\#figure-2}{%
  \includegraphics[width=0.95\textwidth]{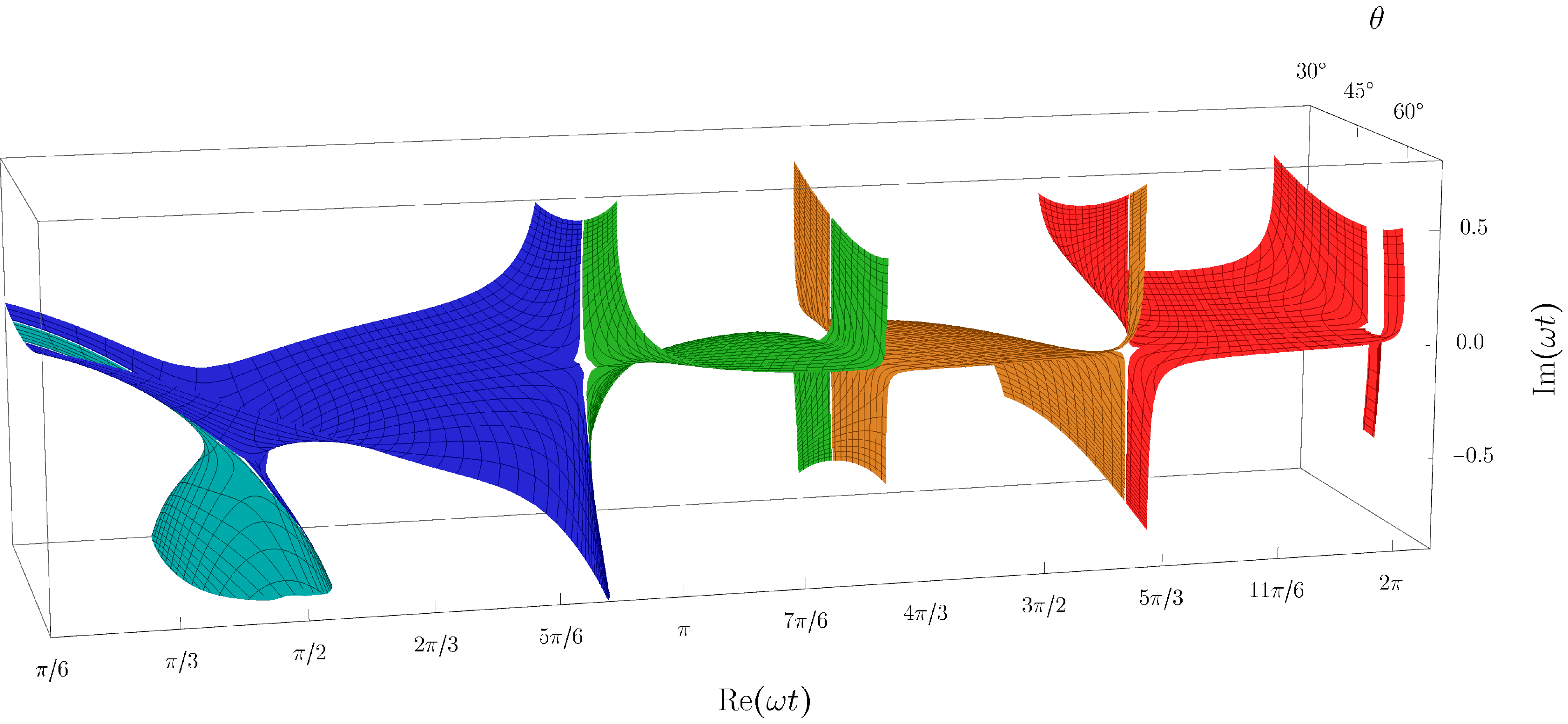} }
\end{tabular}
\caption{
Unified surface formed by the quantum orbits in \reffig{fig-bicircular-transitions}, when the~$\theta$ dependence is unfolded as a third dimension.
(Thus, a cut through the surface at constant $\theta$ will return a panel from \reffig{fig-bicircular-transitions}.)
At the transitions, the horizontal parts of the surface (the plateau harmonics) change from going up to going down and vice versa, but from a global perspective, the saddles form a single surface with a unified topology, and the separation of this surface into individual regions corresponding to the different quantum orbits is somewhat artificial.
This highlights the fact that the various quantum orbits, as inverse images under the multivalued inverse of an analytical function, are simply different branches of a single Riemann surface.
An interactive version and a 3D-printable model of this plot is available at \href{https://imaginarycutoffs.github.io}{imaginary-harmonic-cutoff.github.io}~\cite{SupplementaryMaterial}.
}
\label{fig-topology-3D}
\end{tdfigure*}

\begin{figure*}
{
\setlength{\tabcolsep}{-0.5mm}
\newlength{\figureLspacing}
\setlength{\figureLspacing}{-3.15mm}
\newlength{\figureLspacingwide}
\setlength{\figureLspacingwide}{-1mm}
\begin{tabular}{rl}
\subfloat{\label{fig-bicircular-spectra-a} %
\includegraphics[scale=1]{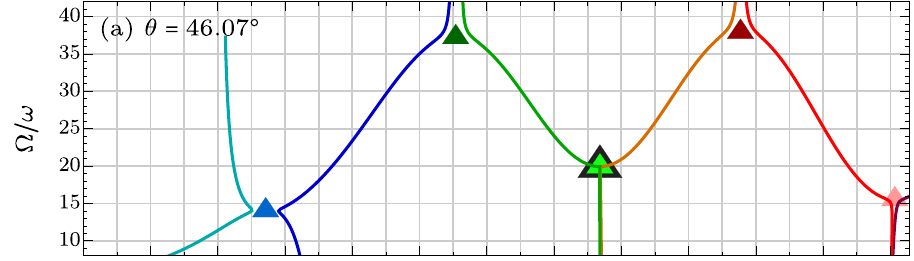} } & 
\subfloat{\label{fig-bicircular-spectra-e} %
\includegraphics[scale=1]{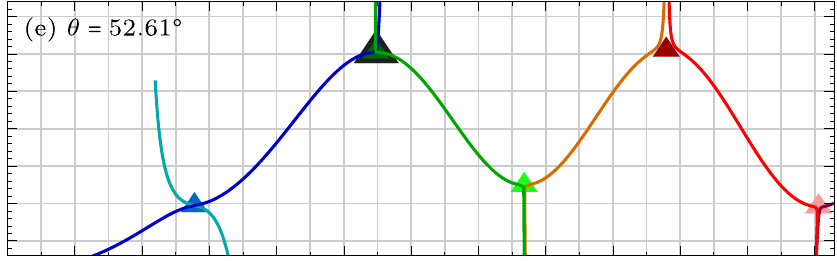} }\\[\figureLspacing]
\subfloat{\label{fig-bicircular-spectra-b} %
\includegraphics[scale=1]{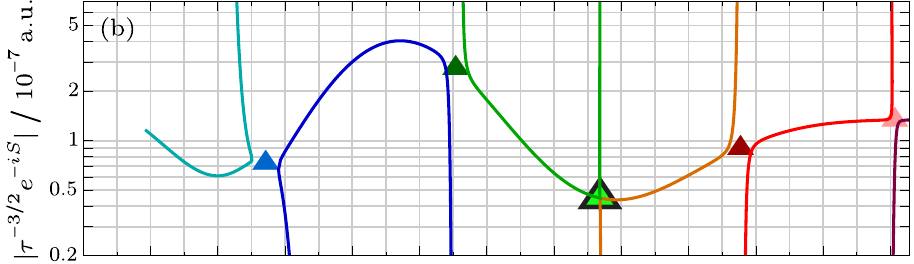} } & 
\subfloat{\label{fig-bicircular-spectra-f} %
\includegraphics[scale=1]{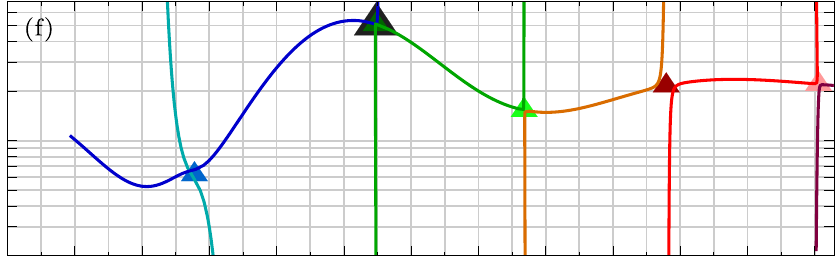} }\\[\figureLspacingwide]
\subfloat{\label{fig-bicircular-spectra-c} %
\includegraphics[scale=1]{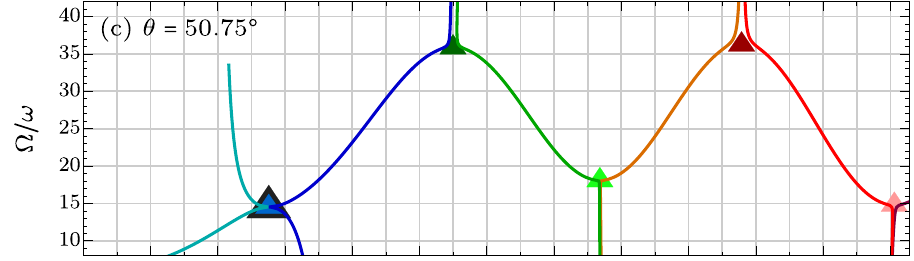} } & 
\subfloat{\label{fig-bicircular-spectra-g} %
\includegraphics[scale=1]{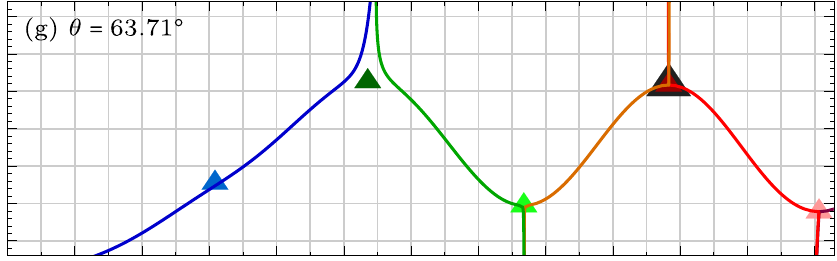} }\\[\figureLspacing]
\subfloat{\label{fig-bicircular-spectra-d} %
\includegraphics[scale=1]{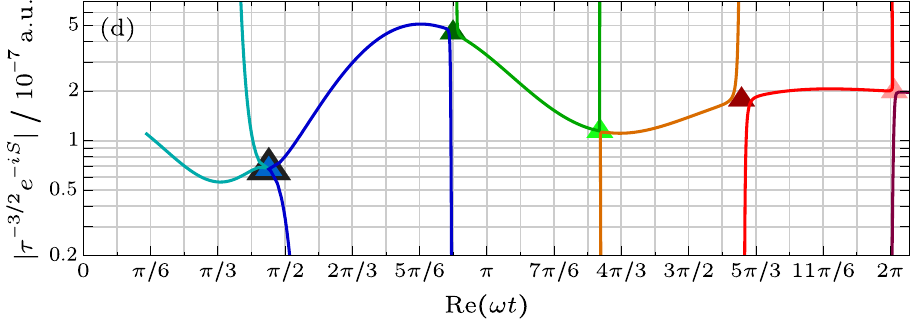} } & 
\subfloat{\label{fig-bicircular-spectra-h} %
\includegraphics[scale=1]{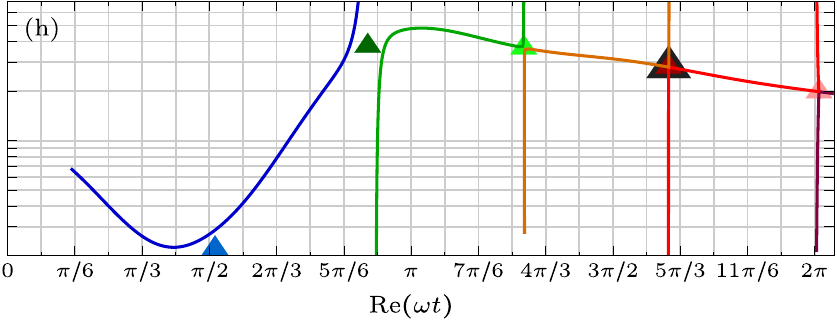} }
\end{tabular}
\vspace{5mm}
}
\caption{
Energy-time relations~(a,\,c,\,e,\,g) and indicative SPA harmonic spectra, including the factors from the action and the wavepacket dilution (b,\,d,\,f,\,h) for four of the quantum-orbit transitions from \reffig{fig-bicircular-transitions}.
For bicircular fields, attention is generally focused on the short quantum orbit (blue curve), since at equal intensities ($\theta=\SI{45}{\degree}$) it dominates the spectrum (as shown in~(b), which closely follows Fig.~8 of \citer{Milosevic2000bicircular}).
However, this is no longer the case at mixing angles with a high intensity for the $2\omega$ field.
At the long-short topological transition (e,\,f), where the two become indistinguishable at the cutoff, the latter is intensified by this effect, 
while at large mixing angles, as shown in (g,\,h), the contribution of the short trajectory plummets.
\vspace{5mm}
}
\label{fig-bicircular-spectra}
\end{figure*}

In other words, the quantum orbits here form a single unified topology, and they should be understood as such.
We show this topology in \reffigtd{fig-topology-3D}, by unfolding the~$\theta$ dependence into a third dimension for the plot: this reveals that the quantum orbits form a single surface, with discontinuous changes in colour where the (apparent) reconnection transitions force us to choose where to split this unified surface into individual components.

The fundamental topological object here, as we explored in depth in Section~\ref{sec-riemann-surface}, is the Riemann surface formed by the saddle points: that is, the surface defined by the equation $\frac{\d S_V}{\d t}(t_s) = \Omega$, encoding the multiple inverses of the action's derivative, whose intersection with the real $\Omega$ axis gives the quantum orbits themselves.
Within that perspective, the difference between the two types of connections between the quantum orbits boils down to what side of the real $\Omega$ axis the branch point (i.e., the harmonic-cutoff time $t_\hc$) falls on.

For fixed field shapes, this branch-cut structure can essentially be ignored, if so desired, since we are only looking for the inverse images for $\Omega$ over the real axis, and those inverse images will normally be well separated.
In \reffigtd{fig-topology-3D}, however, as well as in any other situations where the field shape changes over a control parameter, we see how these changes to the position and shape of the Riemann surface bring these once-separate images into collision (and reconnection) with each other, and this structure can no longer be ignored.

In more general terms, the reconnection phenomenon we have demonstrated in this section entails that, ultimately, attempting to attach labels to individual quantum orbits is fundamentally impossible, since they are essentially just different instances of one and the same object.
Saddle-point tagging schemes of this type, in terms of multi-indices typically notated as $\alpha\beta m$, have been used widely in the literature, both for monochromatic fields~\cite{Chipperfield2007, Milosevic2002, Odzak2005, Chipperfield2005, Milosevic2017chapter} as well as more complex waveforms~\cite{Hasovic2016, Milosevic2019xray, Milovsevic2019quantum, Milosevic2016}. 
The topological features we have explored here imply that these schemes should be regarded as labelling the principal branches of the quantum-orbit Riemann surface.

These schemes can, in principle, be applied to changing driver waveforms with nontrivial connections between the various branches involved, but then those branch changes must be explicitly tracked -- in which case, the branch-point machinery we have developed here (both for finding the branch points $t_\hc$ as well as the parameters where their branch cuts cross the real $\Omega$ axis) is essential.

As a final note here, it is important to remark that, while some of the dynamics we have explored here occurs for quantum orbits with weak or negligible contributions to the spectrum, that is not always the case.
Indeed, as we show in \reffig{fig-bicircular-spectra}, the equal-intensities viewpoint that the spectrum is completely dominated by the short-trajectory contribution breaks down at large mixing angles, where the $2\omega$ contribution is significant, and where many of the topological transitions we have discussed in this section take place.

\section{Discussion and conclusions}
\label{sec-conclusions}
As we have shown, the problem of saddle-point classification can be solved in a robust and flexible way by using the harmonic-cutoff times~$t_\hc$ to locate the center of the missed approach and implement a separatrix that lies between the two quantum orbits that perform an avoided crossing at each cutoff.
These harmonic-cutoff times can be easily found as the zeros of a second-order saddle-point equation.
In a clear sense, the $t_\hc$'s form the centerpiece of the quasiclassical structure at the cutoff, and they are a useful and flexible tool to understand a range of questions about the quantum-orbit dynamics.

The strongest implication of this is that the harmonic-cutoff times can be used to provide a natural identification for the energy of the cutoff, $\Omega_\hc = \frac{\partial S_V}{\partial t}(t_\hc)$. 
This now takes on a complex value, with the imaginary part controlling the strength of quantum-path interference between the quantum orbits that meet at that cutoff, as well as the closeness and direction of the approach between them.

Similarly, our approach provides an efficient method for finding the position of the cutoff as well as the harmonic yield there, which will work uniformly and efficiently for a broad range of optical waveforms for the driving laser. 
This extends to a full estimation of the spectrum, the Harmonic-Cutoff Approximation, which uses only quantities local to $t_\hc$, and which accurately captures the cutoff as well as the qualitative shape of the spectrum down into the middle of the plateau.

On a more abstract note, our search for structures that can be used to classify the solutions of the saddle-point equations into individual quantum orbits yields a fresh perspective on the quasiclassical theory of strong-field phenomena.
The quantum orbits are thus revealed as the $\Im(\frac{\partial S_V}{\partial t})=0$ contour of the time derivative of the action, with the rest of its contour map holding crucial information -- most notably via its saddles, the harmonic-cutoff times $t_\hc$.
Likewise, the usual saddle points are re-understood as individual branches of a larger Riemann surface, which encompasses all of the quantum~orbits.
The branch points that separate these branches are again the harmonic-cutoff times, which can thus be used to watch for changes in the quantum-orbit topology as the driving field's waveform changes.

The location of the cutoff at a zero of the second derivative of the action also admits a rather more pedestrian interpretation, though.
If the emission is modelled using only the classical-trajectory level, then the energy of return $E_\mathrm{return} = \frac{\partial S}{\partial t}$ is a function of the return time, and if we want to find its extrema, then we simply need to solve the equation $\frac{\partial^2 S}{\partial t^2} = 0$.
When the theory is upgraded to the quasiclassical formalism, that equation might seem to lose its meaning, since the action and the trajectories are complex, as would be the solutions to the extremum equation $\frac{\partial^2 S}{\partial t^2} = 0$.
Our solution in this work is entirely in line with the spirit of the quasiclassical formalism for strong-field physics: we solve this extremum equation in the same form it takes classically, allowing for complex values wherever necessary, and then look at the underlying oscillatory integral for the correct interpretation of those complex quantities.

More physically, the key structure at play is the fact that the harmonic cutoff is a caustic: the quantum-orbit analysis of the underlying matter-wave dynamics is exactly analogous to the geometric-optics analysis of wave optics in terms of rays, and in this analogy the harmonic cutoff corresponds to caustics where two families of rays interfere and then meet and fold into each other, giving way to evanescent-wave behaviour analogous to the post-cutoff decay in the harmonic spectrum.
In one dimension, this caustic can be precisely localized to a point, known as the `fold' (more technically, the `bifurcation set'), which marks the transition between the two regimes, and our harmonic-cutoff times are the precise embodiment of that understanding of the caustic.

This view of the harmonic cutoff as a caustic has been used for some time~\cite{Raz2012, Goulielmakis2012, Austin2014}, but recent years have seen a marked increase in interest in that perspective on the dynamics~\cite{Facciala2016, Facciala2018, Hamilton2017, Birulia2019, Uzan2020, Kelvich2017}, as laser sources achieve better control over polychromatic combinations with high relative intensities (which is slated to continue to increase~\cite{Mitra2020}).
This has opened the door to the observation of higher-order catastrophes in strong-field observables, involving higher-dimensional bifurcation sets; however, the theoretical understanding remains somewhat behind, and a quantitative understanding that reflects those structures has not yet followed.
In this work we have precisely pinned down the embodiment of the bifurcation set in the quasiclassical formalism for the `fold' catastrophe, and this can then be used to analyze in detail the more complicated configurations involved in newer experiments.

Similarly, our work suggests that the harmonic-cut\-off times we have discussed here should have analogous structures in ionization experiments (most notably high-order above-threshold ionization~\cite{Figueira2019}), where the experimentally-measurable variable, momentum, has multiple dimensions.
This allows greater freedom to the theory (while at the same time substantially complicating its analysis), and this in turn allows for higher-dimensional singular matter-wave structures to be contained in the experimental results.
The tools we have demonstrated here for understanding the one-dimensional caustic formed at the harmonic cutoff should then provide a useful basis for understanding those configurations.

\acknowledgements
\vspace{-1mm}
We thank Przemyslaw Grzybowski for essential and generous help in understanding the structures presented here.
E.P.~acknowledges Cellex-ICFO-MPQ Fellowship funding.
We acknowledge funding from the Spanish Ministry MINECO (National Plan 15 Grant: FISICATE\-AMO No.\ FIS2016-79508-P, SEVERO OCHOA No.\ SEV-2015-0522, FPI), European Social Fund, Fundació Cellex, Fundació Mir-Puig, Generalitat de Catalunya (AGAUR Grant No.\ 2017 SGR 1341, CERCA/Program), ERC AdG NOQIA, EU FEDER, European Union Regional Development Fund~-- ERDF Operational Program of Catalonia 2014-2020 (Operation Code: IU16-011424), MINECO-EU QUANTERA MAQS (funded by The State Research Agency (AEI) PCI2019-111828-2 / 10.13039/\allowbreak{}501100011033), and the National Science Centre, Poland, Symfonia Grant No.\ 2016/20/W/ST4/00314.

\section*{Author ORCIDs}
\vspace{-1mm}
\begin{itemize}[
  itemsep=-1mm,
  leftmargin=\dimexpr0.3cm+\labelsep\relax,
  label={\smash{\includegraphics[width=8pt]{ORCID-icon.png}}}
  ]
\item Emilio Pisanty:
   \href{https://orcid.org/0000-0003-0598-8524}{0000-0003-0598-8524}
\item Marcelo F. Ciappina:
   \href{https://orcid.org/0000-0002-1123-6460}{0000-0002-1123-6460}
\item Maciej Lewenstein:
   \href{https://orcid.org/0000-0002-0210-7800}{0000-0002-0210-7800}
\end{itemize}

\bibliographystyle{arthur} 
\bibliography{references}{}


\end{document}